\documentclass[12pt]{article}
\pdfoutput=1
\usepackage{soul}
\usepackage{geometry}
\usepackage{amsmath, amssymb, bbm, amsthm, mathabx, bigints}
\usepackage{float}
\usepackage{setspace}
\usepackage[font=small]{caption}
\usepackage[colorlinks,citecolor=DarkBlue,linkcolor=DarkBlue,bookmarks=false,hypertexnames=true]{hyperref} 
\usepackage[svgnames]{xcolor} 
\usepackage{stfloats} 
\usepackage[affil-it]{authblk} 

\usepackage{fancyhdr}
\usepackage[font=footnotesize,labelfont=sc]{caption}
\renewcommand{\thesection}{\Roman{section}} 
\usepackage{titlesec}
\usepackage{lipsum}
\titleformat{\section}{\centering\bfseries}{\thesection. }{0.1em}{}
\titleformat{\subsection}{\centering\normalsize\itshape}{\Alph{subsection}. }{0.1em}{}
\titleformat{\subsubsection}[runin]{\normalsize\itshape}{\thesubsubsection}{1em}{}

\usepackage{graphicx}
\usepackage{grffile}
\usepackage{tablefootnote}
\usepackage{dcolumn}
\usepackage{tabularx}

\usepackage{pdflscape}
\usepackage{pdfpages}

\usepackage{footnote}
\makesavenoteenv{tabular}

\usepackage{csquotes}

\newcommand{\independent}{\perp \!\!\! \perp} 

\usepackage{natbib}

\renewenvironment{abstract}
 {\small
  \begin{center}
  \bfseries \abstractname\vspace{-.5em}\vspace{0pt}
  \end{center}
  \list{}{%
    \setlength{\leftmargin}{21mm}
    \setlength{\rightmargin}{\leftmargin}%
  }%
  \item\relax}
 {\endlist}

\begin{document}

\makeatletter
\patchcmd{\@maketitle}{\LARGE \@title}{\fontsize{12}{12}\selectfont\@title}{}{}
\newcommand\customsize{\@setfontsize\customsize{8.5}{11}}
\makeatother

\renewcommand\Authfont{\fontsize{12}{12}\selectfont} 
\renewcommand{\abstractname}{}    

\newcommand\blfootnote[1]{%
  \begingroup
  \renewcommand\thefootnote{}\footnote{#1}%
  \addtocounter{footnote}{-1}%
  \endgroup
}

\font\myfont=cmr12 at 16pt
\title{{\myfont Can the Replication Rate Tell Us About Publication Bias?}}
\author[1]{\textsc{By Patrick Vu}\footnote{\textit{This version:} \today. Brown University. patrick\_vu@brown.edu. I am especially grateful for the feedback, advice, and encouragement of Jonathan Roth. For helpful comments, suggestions and conversations, I thank Johannes Abeler, Daniel Bj\"{o}rkegren, Pedro Dal B\'{o}, Peter Hull, Toru Kitagawa, and Soonwoo Kwon, as well as seminar participants at Brown University.}}
\date{\vspace{-10ex}}

\maketitle
\begin{abstract}
\textit{A leading explanation for widespread replication failures is publication bias. I show in a simple model of selective publication that, contrary to common perceptions, the replication rate is unaffected by the suppression of insignificant results in the publication process. I show further that the expected replication rate falls below intended power owing to issues with common power calculations. I empirically calibrate a model of selective publication and find that power issues alone can explain the entirety of the gap between the replication rate and intended power in experimental economics. In psychology, these issues explain two-thirds of the gap.}
\end{abstract}

A key stylized fact about the replication crisis in the social and life sciences is that the share of statistically significant results that are replicated is relatively low. In a replication of 18 experimental economics studies, \citet{Camerer2016} found that 61\% percent of significant results were replicated with the same sign and significance. In psychology, \citet{OpenScience2015} found that only 36\% of significant results were successfully replicated. In both cases, the target for the replication rate, which we refer to as intended power, was 92\%. Explaining the source of this gap has become a topic of intense interest. Prominent explanations include the suppression of null results in the publication process, differences in design across original studies and replications, p-hacking, manipulation, and measurement error in small samples.\footnote{References on each of these explanations include: (i) Publication bias: \citet{OpenScience2015, Maxwell2015, Camerer2016, Anderson2017, Camerer2018, Stanley2018}; (ii) Treatment effect heterogeneity: \citet{Higgins2002, Cesario2014, Simons2014, Stanley2018, Bryan2019}; (iii) p-hacking and manipulation: \citet{Ioannidis2005, Ioannidis2008, Simonsohn2014, Brodeur2016}; (iv) Measurement error: \citet{Gelman2014, Loken2017, Gelman2018}.} 

In this article, I argue that publication bias is unlikely to explain low replication rates, and then show that replication rate gaps should be expected even in the absence of the other issues cited above. The argument is summarized in two theoretical results. First, I show in a simple model of selective publication that the replication rate does not depend on the probability with which insignificant results are published. This runs counter to the common perception that suppression of null results in the publication process is a key contributor to low replication rates. While this result is somewhat counter-intuitive, its explanation is simple. The replication rate is defined as the share of \textit{significant} results that are replicated with significance and the same sign. Since the replication rate definition does not depend on insignificant results, it is unaffected by their prevalence in published work. It is important to note that this result does not depend on whether or not insignificant findings are chosen for replication; even when they are, the replication rate calculation does not include them (e.g. \citet{OpenScience2015}). The replication rate is thus ill-suited to uncovering the most salient form of publication bias.

The second theoretical result shows that low replication rates should nevertheless be expected. This occurs because of three issues with common power calculations. First, original estimates included in the replication rate are not a random sample of published findings, but instead a selected sample of significant findings. It is well known that samples selected on extreme characteristics (e.g. height, test scores, statistical significance) will regress to the mean in repeated samples \citep{Galton1886, Hotelling1933, Barnett2004, Kahneman2011}. In replication settings, this means that the sample of significant original estimates used to calculate the replication rate are mechanically inflated in expectation, and that replication estimates will regress to the mean. Power calculations in replications calibrated to detect inflated original estimates will therefore be systematically underpowered for recovering smaller true effects, leading to low replication rates. Again, whether or not insignificant findings are chosen for replication has no bearing on this conclusion. A second issue is that common power calculations do not account for the non-linearity of the power function. I show that the expected replication rate falls below intended power even in the optimistic scenario where original estimates are unbiased for true effects, all studies are published irrespective of significance, and replications are a random sample of the published literature. Finally, common power calculations lead to very low replication probabilities when original estimates are the opposite sign of the true effect, which occurs with positive probability due to random sampling variation. This is because a `successful' replication in this case relies on the highly unlikely outcome that a replication reproduces a statistically significant result with the `wrong' sign.

These three issues with common power calculations imply that the expected replication rate falls below intended power. Importantly, this result holds even under `ideal' conditions without publication bias, heterogeneity in true effects, p-hacking, manipulation, and measurement error. This suggests that intended power targets in large-scale replication studies do not provide a meaningful benchmark against which to judge replication rates observed in practice; low replication rates are what we should expect.

To what extent can power issues alone explain observed replication rate gaps? To answer this question, I estimate a model of selective publication based on \textcolor{DarkBlue}{Andrews and Kasy} (\citeyear{Andrews2019}) for large-scale replication studies in experimental economics \citep{Camerer2016} and psychology \citep{OpenScience2015}. Estimation uses data from original studies but not replication studies. This allows us to make out-of-sample `predictions' of the fraction of results that replicate, using the power calculations actually implemented in replications. The empirical exercise asks, in effect, whether observed replication rate gaps could have been predicted by issues with common power calculations alone, before the replication studies themselves were actually undertaken.

The estimated model accounts for the entirety of the replication rate gap in experimental economics. It predicts a replication rate of 60\%, compared to the 61\% actually observed. In psychology, the model predicts a replication rate of 54\%. This is well below mean intended power of 92\%, but still above the observed replication rate of 36\%. Thus the model can account for two-thirds of the gap. This discrepancy suggests that factors not included in the model -- for example, heterogeneity in true effects, p-hacking, measurement error -- may be important in psychology.

To be clear, the results in this article do not suggest that there is no publication bias. The prevalence of publication bias and its distortions are well documented \citep{Ioannidis2008, Franco2014, Gelman2014, Landis2014, Mervis2014, Gelman2018, Andrews2019, Abadie2020}. Responses to mitigate these distortions include results-blind peer review \citep{Chambers2013, Foster2019}, journals dedicated to publishing insignificant findings\footnote{Examples include: \textit{Positively Negative (PLOS One); Journal of Negative Results in Biomedicine; Journal of Articles in Support of the Null Hypothesis; Journal of Negative Results - Ecology and Evolutionary Biology}.}, and even cash incentives for publishing null findings (\textcolor{DarkBlue}{Nature} \citeyear{Nature2020}). The results in this article suggest that the replication rate is a poor metric to gauge whether such reforms are successful in reducing publication bias.

The final section of this article conducts counterfactual exercises based on the estimated model to evaluate the extent to which the replication rate and other measures of publication bias would respond to changes in the probability of publishing insignificant results. In line with the theoretical results of the model, the replication rate is completely unresponsive to changes in the probability of publishing insignificant results. By contrast, other measures such as mean bias and coverage are more responsive. This suggests that measures other than the replication rate are better-suited to evaluate efforts to reduce publication bias.

This article contributes to the large literature on metascience and publication bias \citep{Card1995, Ioannidis2005, Rothstein2006, Gorroochurn2007, Ioannidis2008, Button2013, Franco2014, Gelman2014, Landis2014, Mervis2014, Maxwell2015, Anderson2017, Stanley2017, Stanley2018, Gelman2018, Klein2018, Miguel2018, Shrout2018, Amrhein2019, AmrheinNature2019, Tackett2019, Andrews2019, Miguel2019, Kasy2022}. It is not the first to question the replication rate. \citet{AmrheinNature2019} criticize the replication rate because it emphasizes statistical significance over scientific significance. This can lead to incongruous conclusions. For example, two studies with identical point estimates, but where one is statistically significant and the other is not due to  sample size differences, will be counted as `inconsistent' under the current definition of the replication rate. Separately, \citet{Andrews2019} and \citet{Kasy2021} provide stylized examples showing that almost any replication probability is possible when there is no publication bias. This article contributes to these criticisms. It establishes formally that we should expect the replication rate to fall below its nominal target owing to issues with common power calculations, and shows empirically that these issues can explain the entirety of the replication rate gap in experimental economics. It differs from previous literature by attributing inflated effect sizes to the conditioning strategy inherent in the replication rate statistic, rather than publication bias. While suppression of null results leads to inflated effect sizes in the set of all published results, this article argues, by contrast, that it should have no direct impact on the set of significant results used to calculate the replication rate. Empirically, this article builds on \citet{Anderson2017}, which calculates replication rates using fully simulated data. I calculate the replication rate using a model empirically calibrated on data from replication studies. This allows for a comparison between model-based predictions and observed replication rates. 

\section{Simple Example}
A simple example illustrates the key ideas. Note that publication bias in this article refers collectively to journals' decisions on whether to publish studies based on their results and researchers' decisions on whether to write up and submit findings based on the results.

Consider research on the impact of a new drug on health outcomes. Assume the true treatment effect is $\theta=2.5$. Researchers conduct a large number of independent studies, each producing an estimated effect size $X^*$ drawn from a $N(2.5,1)$ distribution. However, only a subset are published because of publication bias. Denote published studies as $X$, which come from the distribution of $X^*$ conditional on publication. Now suppose a large-scale replication is conducted. Replication estimates $X_r$ are drawn from a $N(2.5,\sigma_r(X)^2)$ distribution, where replication standard errors $\sigma_r(X)$ are calculated to detect the original estimates $X$ with 90\% power. This method of calculating power is perhaps the most common approach in replications, and is implemented in the two applications considered in this article. The question we are interested in answering is: What is the replication rate under different standards for publishing statistically significant and insignificant results? 

We consider three vastly different publication regimes and show that all produce exactly the same replication rate. First, the no publication bias regime, where all results are published irrespective of their statistical significance. Second, a regime that publishes all significant results and censors all insignificant results. Third, a regime where insignificant results are five times \textit{more} likely to be published than significant results. The first row of Figure \ref{fig:proposition_intuition} shows the relative publication probabilities for each of these regimes at different $t$-ratios. The second row shows the implied distribution of published estimates $X$. When there is no publication bias, published estimates are drawn from a normal distribution. However, selection on significance reweights different regions of the distribution. When insignificant findings are never published, the insignificant region of the density shrinks to zero; when insignificant results are favoured over significant results, the density over the insignificant region is magnified, and the `significant' tails shrink. The critical point is what occurs when we consider the subset of published results that are statistically significant. This is the set of results included in the replication rate. After conditioning on significant results, all three publication regimes have identical distributions (row 3 in Figure \ref{fig:proposition_intuition}). This implies that the replication rate -- that is, the share of significant findings with the same sign and significance in replications -- must be the same under all publication regimes.

\begin{figure} [tp]
	\centering
\includegraphics[width=1\textwidth]{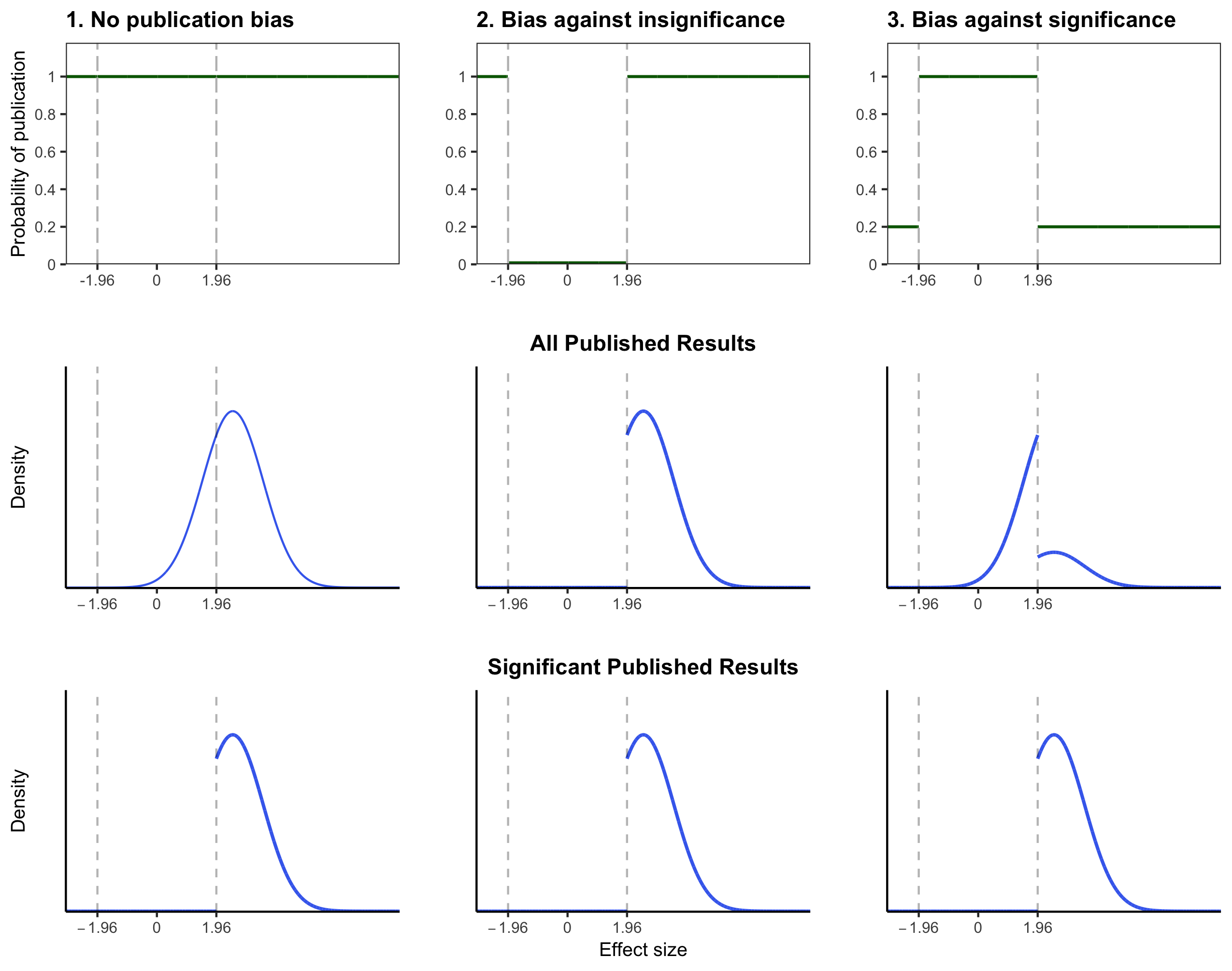} 
\end{figure}

\begin{table}[bp] \centering 
\customsize
\begin{tabularx}{\textwidth}{lcccc} \\
[-1.8ex]\hline 
\hline \\[-1.8ex] 
& & Regime 1 & Regime 2  & Regime 3  \\
\hline
\underline{\textit{All published results}} & &  &   &   \\
Expectation of original estimates & $\mathbbm{E}(X)$ & $2.50$ & $2.99$ & $1.87$ \\ 
Bias of original estimates & $\mathbbm{E}(X) - \theta$ & $0.00$ & $0.49$ & $-0.63$ \\ 
& & & & \\
\underline{\textit{Significant published results}} & &  &   &   \\
Expectation of original estimates & $\mathbbm{E}(X | X \geq 1.96)$ & $2.99$ & $2.99$ & $2.99$ \\ 
Expectation of replication estimates & $\mathbbm{E}(X_r | X \geq 1.96) = \theta$ & $2.50$ & $2.50$ & $2.50$ \\ 
Replication probability (90\% target) & $\mathbbm{P}(|X_r| \geq 1.96 \sigma_r, \text{sgn}(X_r) = \text{sgn}(X) | \theta = 2.5, \sigma_r$) & $0.77$ & $0.77$ & $0.77$ \\ 
\hline 
\hline \\[-1.8ex]
\end{tabularx} 
  \captionof{figure}{\textsc{Publication Regimes, Distributions of Published Results, and the Replication Rate}}
  \caption*{\textit{Notes:} Published estimates $X$ are assumed to be drawn from a normal distribution centered at $\theta=2.5$ with standard error $\sigma=1$, which may be reweighted based on the publication bias function. In Regime 1, all results are published. In Regime 2, only statistically significant results are published. In Regime 3, insignificant results are five times more likely to be published than significant results. Replication estimates $X_r$ are drawn from a $N(2.5, \sigma_r^2)$ distribution, where $\sigma_r$ is set to detect the original estimates with 90\% power using the common power rule. Statistics are based on $10^6$ simulation draws.}
  \label{fig:proposition_intuition}
\end{table}

There are three main takeaways. Each is highlighted by the descriptive statistics at the bottom of Figure \ref{fig:proposition_intuition}. First, the replication rate does not depend on the probability of publishing insignificant results. Publication bias against insignificant findings is therefore unlikely to explain the low replication rates observed in practice. A caveat is that publication bias against null findings may incentivize researchers to manipulate results to obtain significance. This analysis shows that the replication rate will fall short of intended power even in the absence of such manipulation. The empirical results show that a model without manipulation can account fully for the observed replication rate gap in experimental economics.

The second takeaway is that conditioning published studies on significance induces upward bias in original estimates. The replication rate definition imposes this conditioning such that the statistic itself induces inflationary bias. Unbiased replication estimates regress to the mean. This is a consequence of the replication rate definition itself and is the same across all publication regimes. 

Third, the replication rate is below 90\% intended power in all three regimes. Low replication rates are the result of three issues with common power calculations. First, power calculations are calibrated to detect original effect sizes, which are inflated because the replication rate conditions on significance. True effects will be mechanically smaller on average, and thus the replication rate will fall below intended power in expectation. Second, common power calculations do not account for the non-linearity of the power function. Third, common power calculations deliver very low replication probabilities when original estimates are the opposite sign of the true effect, which occurs with non-zero probability. Again, low replication rates in this example are not affected by the extent to which null findings are, or are not, published; moreover, these conclusions hold even in the absence of heterogeneity in true effects, p-hacking and manipulation.

\section{General Case}\label{sec:theory}
Conclusions in the simple example hold more generally. This section formalizes these ideas in a general setting, building on the model of selective publication in \citet{Andrews2019}. 

\subsection{Model of Large-Scale Replication Studies}
Suppose a large-scale replication study is conducted and we observe the estimated effect sizes and standard errors for original studies and their replications. The data-generating process of these studies is modelled as a truncated sampling process. The model is presented here in general form, while the empirical applications make distributional and functional form assumptions. Upper case letters denote random variables, lower case letters realizations. Latent studies (published or unpublished) have a superscript * and published studies have no superscript. The model has five stages:

\begin{enumerate}
    \item \textbf{Draw a population parameter and standard error:} Draw a research question with population parameter ($\Theta^*$) and standard error ($\Sigma^*$):
    $$
    (\Theta^*, \Sigma^*) \sim \mu_{\Theta, \Sigma}
    $$
    where $\mu_{\Theta, \Sigma}$ is the joint distribution of these random variables.
    \item \textbf{Estimate the effect:} Draw an estimated effect from a normal distribution with parameters from Stage 1: 
    $$
   X^* | \Theta^*, \Sigma^* \sim N(\Theta^{*}, \Sigma^{*2})
    $$
    \item \textbf{Publication selection:} Selective publication is modelled by the function $p()$, which returns the probability of publication for any given $t$-ratio. Let $D$ be a Bernoulli random variable equal to 1 if the study is published and 0 otherwise, where\footnote{Notation distinguishing the publication probability function $p()$ over significant and insignificant regions is convenient for presenting the formal results; the cutoff at 1.96 corresponds to the threshold over which results are included in the replication rate.}
    $$
    \mathbbm{P}(D = 1 | X^*/\Sigma^*) =     \begin{cases} 
          p_{sig}(X^*/\Sigma^*) & \text{if } 	\big|\frac{X^*}{\Sigma^*}\big| \geq 1.96 \\
          p_{insig}(X^*/\Sigma^*) & \text{if } \big|\frac{X^*}{\Sigma^*}\big| < 1.96  \\
    \end{cases}
    $$
\end{enumerate}

\begin{enumerate}
    \setcounter{enumi}{3}
        \item \textbf{Replication selection:} Replications are sampled from published studies $(X, \Sigma, \Theta)$; that is, the distribution of latent studies $(X^*, \Sigma^*, \Theta^*)$ conditional on publication $(D=1)$. Replication selection is modelled by the function $r()$, which returns the probability of being chosen for replication for any given $t$-ratio. Let $R$ be a Bernoulli random variable equal to 1 if the study is chosen for replication and 0 otherwise, where
    $$
    \mathbbm{P}(R = 1 | X/\Sigma) =     \begin{cases} 
          r_{sig}(X/\Sigma) & \text{if } 	\big|\frac{X}{\Sigma}\big| \geq 1.96 \\
          r_{insig}(X/\Sigma) & \text{if } \big|\frac{X}{\Sigma}\big| < 1.96  \\
    \end{cases}
    $$
    \item \textbf{Replication:} For results chosen for replication, a replication draw is made with
    $$
        X_r | X, \Sigma, \Theta, \sigma_r^2(X, \Sigma, \beta^n), D = 1, R = 1 \sim N\Big(\Theta,\sigma_r^2(X, \Sigma, \beta^n) \Big)
    $$
    \noindent
    where replication standard errors $\sigma_r(X, \Sigma, \beta^n)$ are chosen by replicators as a function of the original estimate, standard error, and intended statistical power $1-\beta^n$.
\end{enumerate}

\sloppy

We observe i.i.d draws of $\big(X, \Sigma, X_r, \sigma_r(X, \Sigma, \beta^n) \big)$ from the conditional distribution of $\big(X^*, \Sigma^*, X_r, \sigma_r(X^*, \Sigma^*, \beta^n) \big)$ given $D=1$ and $R=1$. The \citet{Andrews2019} model consists of the first three steps, which are used to identify and estimate $p()$. Subsequent replication steps are introduced to analyze the replication rate.

Step 4 models the replication selection mechanism. This differs across replication studies. For the theory, we assume that the set of significant results chosen for replication is a random sample from published, significant results; selection of insignificant findings for replication, $r_{insig}()$, has no impact on the conclusions and can take any form.

Step 5 models how replicators set statistical power. This is a critical factor in determining replication probabilities. Note that replication estimates are assumed to be unbiased estimates of the true effect and generated from exact replications (i.e. a single true effect $\Theta$ across original and replication studies).

In what follows, we normalize $\theta$ to be positive and assume that the distribution of true effects, $\mu_\Theta$, has support on an open set on the positive real line.\footnote{Large-scale replications include studies that examine different questions and outcomes. Normalizing true effects to be positive is justified because relative signs across studies are arbitrary.} The joint probability of publication and being chosen for replication is identified up to scale. Proofs make use of properties of replication probability function presented in Appendix A. Proposition proofs not included in the text are in Appendix B.

\subsection{The Replication Rate and Publication Bias} We begin by defining the replication probability of a single study and then use this to define the expected replication probability over multiple studies. \\

\noindent
\textbf{Definition 1} (Replication probability of a single study). \textit{The replication probability of a published study $(X, \Sigma, \Theta)$ chosen for replication $(R=1)$ is}
\small
\begin{equation}
    RP\Big( X, \Theta, \sigma_r(X, \Sigma, \beta^n) \Big) =     \mathbbm{P} \Bigg( \frac{| {X}_r | }{\sigma_r(X, \Sigma, \beta^n)} \geq 1.96 , \text{sign}(X_r) = \text{sign}(X)
    \Big|
    X, \Theta, \sigma_r(X, \Sigma, \beta^n)
    \Bigg)
\end{equation}
\normalsize

This definition captures the dual requirement that the replication estimate is statistically significant with the same sign as the original study. The replication rate is an aggregate statistic based the fraction of `successful' replications across multiple original studies. The population analogue of the replication rate is defined next: \\

\noindent
\textbf{Definition 2} (Expected replication probability). \textit{The expected replication probability is defined over published studies $(X, \Sigma, \Theta)$ which are chosen for replication $(R=1)$ and statistically significant $(|X/\Sigma| \geq 1.96)$. It is equal to}
\footnotesize
$$
    \mathbbm{E}\Big[
    RP\big( 
    X, \Theta, \sigma_r(X, \Sigma, \beta^n) 
    \big) 
    \big| 
     R=1,  |X/\Sigma| \geq 1.96 \Big]
$$
\begin{equation}
    =
    \int
    RP\Big( x, \theta, \sigma_r(x, \sigma, \beta^n) \Big)
    f_{X^*, \Sigma^*, \Theta^* | D, R, S_X^*}\big(x, \sigma, \theta | D=1, R=1, 1.96 \leq |X^*/\Sigma^*| \big)dx d\sigma d\theta
    \label{def:2}
\end{equation}
\normalsize

This definition highlights an important distinction between being chosen for replication and being included in the replication rate calculation: while insignificant results may be replicated, they are not, by definition, included in the replication rate. This is the definition of the replication rate reported in most large-scale replication studies \citep{Klein2014, OpenScience2015, Camerer2016, Camerer2018, Klein2018}. With this, we can state our first main result. \\

\newpage
\noindent
\textbf{Proposition 1} (The replication rate does not depend on publication bias for null results). \textit{The expected replication probability depends on the probability of publishing significant results, $p_{sig}()$, and does not depend on the probability of publishing insignificant results, $p_{insig}()$.}

\begin{proof}
The expected replication probability can be written as

\footnotesize
$$
    \mathbbm{E}\Big[
    RP\big( 
    X, \Theta, \sigma_r(X, \Sigma, \beta^n) 
    \big) 
    \big| 
     R=1,  |X/\Sigma| \geq 1.96 \Big]
$$
\begin{equation}
    =
    \bigintssss
    \Bigg( 1 -  \Phi \bigg( 1.96 - \text{sign}(x) \frac{\theta}{\sigma_r(x, \sigma, \beta^n) } \bigg) \Bigg)
    f_{X^*, \Sigma^*, \Theta^* | D, R, S_X^*}\big(x, \sigma, \theta | D=1, R=1, | X^*/\Sigma^*| \geq 1.96 \big)dx d\sigma d\theta
    \label{eq:obs1}
\end{equation}
\normalsize

\noindent
where $f_{X^*, \Sigma^*, \Theta^* | D, R, S_X^*}$ is the joint distribution of original studies conditional on being published, chosen for replication, and statistically significant at the 5\% level; the integrand is derived in Lemma 2.1 in Appendix A; and $\text{sign}(x) = 1$ if $x>0$ and $-1$ otherwise. The conditional distribution does not depend on the probability of publishing insignificant findings. To see this, apply Bayes rule twice to get

\footnotesize
$$
f_{X^*, \Sigma^*, \Theta^* | D, R, S_X^*}\big(x, \sigma, \theta | D=1, R=1, | X^*/\Sigma^*| \geq 1.96 \big)
$$
$$
    = \frac{\mathbbm{P} \Big(D=1 \big| X^* = x, \Sigma^* = \sigma, \Theta^* = \theta, R=1, | X^*/\Sigma^*| \geq 1.96 \Big)}
    {\mathbbm{P}\Big(D=1 \big| R=1, | X^*/\Sigma^*| \geq 1.96 \Big)}
    \times
    \frac{\mathbbm{P} \Big(R=1 \big| X^* = x, \Sigma^* = \sigma, \Theta^* = \theta, | X^*/\Sigma^*| \geq 1.96 \Big)}
    {\mathbbm{P}\Big(R=1 \big| | X^*/\Sigma^*| \geq 1.96 \Big)}
$$
$$
    \times
    f_{X^*, \Theta, \Sigma^* | S_X^*} \Big(x, \theta, \sigma \big| | X^*/\Sigma^*| \geq 1.96 \Big)
$$
\begin{equation}
    = 
    \frac{p_{sig}(x/\sigma) }
    {\mathbbm{E} \big( p_{sig}(X^*/\Sigma^*) \big| 1.96 \leq | X^*/\Sigma^* | \big) }
    f_{,X^*, \Sigma^*, \Theta^* | S_X^*} \Big(\theta, x, \sigma \big| | X^*/\Sigma^*| \geq 1.96 \Big) 
    \label{eq:obs1proof}
\end{equation}
\normalsize

\noindent
In the second line, the second term in the product equals one because replication selection for significant results is assumed to be random. In the last line, the ratio in the first term in the product includes only $p_{sig}()$; the denominator does not condition on $R$ because replication selection is random for significant findings. The second term in the product is the joint density of latent variables conditional on significance, which is not affected by publication bias.
\end{proof}

Proposition 1 is somewhat counter-intuitive. It follows because conditioning out statistically insignificant results makes the replication rate uninformative about the degree to which such results are or are not published. The replication rate instead depends on replication power, the distribution of latent original studies, and the relative probability of publication when the absolute value of the $t$-ratio is greater than 1.96, $p_{sig}()$. Appendix C provides an example showing how the replication rate varies as we change $p_{sig}()$.

A caveat is that the model assumes a fixed distribution of latent studies, whereas in practice it may be endogenous. For example, changes in the publication of insignificant results could alter the behaviour of researchers, by changing the likelihood that they engage in specification searching or manipulation \citep{Simonsohn2014, Brodeur2016}. Incorporating such behavior in the model would allow the publication probability function $p_{insig}()$ to affect either the joint distribution $\mu_{\Theta, \Sigma}$ (researchers changing the questions they ask); the interdependence of draws from $\mu_{\Theta, \Sigma}$ within a study (multiple hypothesis testing and specification searching); or the distribution of the estimated effect (manipulation of findings), e.g. if $X^* \in (1.96\sigma-\epsilon, 1.96\sigma)$ with $\epsilon>0$, then with some probability the researcher reports $X^* + \epsilon$. 

\subsection{Common Power Calculations and Low Replication Rates}\label{sec:pitfalls}

The remaining propositions show that issues with common power calculations can lead to replication rates that fall consistently below intended power. \\

\noindent
\textbf{Definition 3} (Common power rule). \textit{The common power rule to detect original effect size $x$ with intended power $1-\beta^n$ sets the replication standard error to}

\begin{equation}
    \sigma_r(x, \beta^n) = \frac{|x|}{1.96 - \Phi^{-1} (\beta^n)}
\end{equation} 

\textit{This is equivalent to setting the replication sample size to $N \times \big[ \sigma (1.96 - \Phi^{-1}(\beta^n))/x)\big]$, where $N$ and $\sigma$ are the original study's sample size and standard deviation, respectively.} \\

\noindent
\textbf{Lemma 1.} (Justification of the common power rule). \textit{Consider a published study $(x, \sigma, \theta)$. If $x = \theta$ and a replication uses the common power rule to detect the original effect with intended power $1-\beta^n$, then}

\begin{equation}
    RP\Big( \theta, \theta, \sigma_r(\theta, \beta^n) \Big) = 1 - \beta^n
    \label{eq:lemma1}
\end{equation}

\begin{proof}
Substitute the common power rule in the replication probability function. If $x=\theta$, then
\small
\begin{equation}
    RP\big( \theta, \theta, \sigma_r(\theta, \beta^n)  \big)
    = 1 -  \Phi \bigg( 1.96 - \text{sign}(\theta) \frac{\theta}{\sigma_r(\theta, \beta^n)} \bigg) =
    1 -  \Phi \bigg( 1.96 - \frac{\theta}{\theta} \big(1.96 - \Phi^{-1}(\beta^n)\big) \bigg) = 1 - \beta^n
\end{equation}
\normalsize
\end{proof}

Lemma 1 provides the justification for the common power rule. Its reasoning is as follows. Replication probabilities depend crucially on the unobserved true effect $\theta$. If there are no issues with the original study, then its effect size $x$ should be a reasonable proxy for the true effect $\theta$ for setting statistical power. Extending this idea to multiple studies suggests that the replication rate should be close to intended power $1 - \beta^n$. In practice, it consistently falls below this benchmark \citep{OpenScience2015, Camerer2016, Camerer2018, Klein2018}. This is commonly interpreted as an indicator of problems with original studies, replication studies, or both. 

The remaining propositions show that this need not be the case. Three issues with the common power rule imply that the expected replication rate falls short of intended power even in the absence of any such problems. For these propositions, I assume that $p_{sig}()$ is symmetric about zero, non-zero, bounded, differentiable and weakly increasing in absolute value; no restrictions are placed on $p_{insig}()$. The first issue is based on the following result:
\\

\noindent
\textbf{Proposition 2} (Regression to the mean in replications). \textit{Published original estimates $X$ and corresponding replication estimates $X_r$ satisfy}

\begin{equation}
    \mathbbm{E}\big[X \big| \Theta = \theta, |X/\Sigma| \geq 1.96 \big]
    >\theta = \mathbbm{E} \big[X_r| \Theta = \theta \big]
    \label{eq:reg_to_mean}
\end{equation} \\

The replication rate conditions on statistical significance. Thus, upwardly biased effect sizes in the set of results included in the replication rate is an inevitable consequence of the replication rate definition itself. This holds irrespective of publication bias on null results. Regression to the mean in (unbiased) replication attempts is to be expected. This statistical fact invalidates the assumption underpinning the justification of the common power rule in Lemma 1. In general, a significant original estimate $X$ is not an unbiased proxy for the unobserved true effect $\Theta$.

A second issue is that the common power rule does not account for the fact that the replication probability is a non-linear function of original estimates $X$. This implies that even for unbiased original estimates, the expected replication probability does not in general equal the replication probability evaluated at true effect (i.e. for $X \sim N(\theta, \sigma^2)$, in general: $\mathbbm{E}[RP(X, \Theta, \sigma_r(X, \beta^n) ) | \Theta = \theta] \neq RP(\theta, \theta, \sigma_r(\theta, \beta^n) ) = 1 - \beta^n$). Third, random sampling variation implies that original estimates will occasionally have the `wrong' sign. In this case, the replication probability is bounded above by 0.025 since $X<0$ implies $RP\big(X, \Theta, \sigma_r(X, \beta^n)\big) = \Phi \big(-1.96 - \frac{\Theta}{\sigma_r(X, \beta^n)} \big) < 0.025$. The likelihood that original estimates have the opposite sign is higher in settings with low power and small true effects \citep{Gelman2014, Stanley2017}.\footnote{ Note that both the second and third issue arise because the justification of the common power rule does not account for the fact that original estimates $X$ are random variables.} These three issues lead to the following: \\

\newpage
\noindent
\textbf{Proposition 3} (The common power rule implies low expected replication rates). \textit{Suppose replication standard errors are set by the common power rule to detect original estimates with intended power $1-\beta^n \geq 0.8314$. Then}

\begin{equation}
    \mathbbm{E}\Big[
    RP\big( 
    X, \Theta, \sigma_r(X, \Sigma, \beta^n) 
    \big) 
    \big| 
     R=1, |X/\Sigma| \geq 1.96  \Big]
    <
    1- \beta^n
\end{equation} \\

Proposition 3 is intuitive. If replication estimates regress to the mean, then common power calculations calibrated to detect inflated original effects will tend to be underpowered for recovering smaller true effects. For original results with the `wrong' sign, `successful' replications are especially unlikely because they require that authors reproduce a result with the opposite sign of the true effect. Low replication rates observed in practice should therefore come as no surprise. The requirement that intended power satisfies $1-\beta^n \geq 0.8314$ relates to the non-linearity of the replication probability function, and is met in both empirical applications we consider, where mean intended power is 0.92 \citep{OpenScience2015, Camerer2016}. 

Several points are worth noting. First, Proposition 3 holds even under `ideal' conditions of no publication bias, no p-hacking or manipulation, replications with identical designs and comparable samples (i.e. no heterogeneity in true effects), no measurement error, and random sampling in replication selection. The expected replication probability still falls below intended power in this case, and thus points to fundamental difficulties in interpreting replication rate gaps observed in large-scale replication studies.

Second, extending the replication rate definition to include insignificant results is unlikely to be a meaningful corrective. Given the problems that stem from conditioning on statistical significance, a natural proposal might be to calculate the replication rate over a random sample of all results, irrespective of their significance. There are at least two problems with this. First, even in the optimistic scenario in which all results are published, unbiased, and included in the replication rate, the expected replication probability still falls short of intended power because of the other two issues with common power calculations.\footnote{This result is presented in Appendix B as Lemma 4 and is used as an intermediate step in the proof of Proposition 3.} Second, and perhaps more fundamentally, the replication rate is not conceptually coherent for research findings where the null hypothesis is true. If the true effect is zero, then replication `success' under the current definition requires that the replication attempt is statistically significant.\footnote{Appendix D considers an alternative proposal for extending the replication rate definition to obtaining insignificance in replications when an original finding is insignificant. In simulations, this generalized replication rate remains below intended power under the common power rule.}

\section{Empirical Applications}
To what extent can Proposition 3 explain low replication rates observed in practice? To answer this question, I estimate the model in Section \ref{sec:theory} for two large-scale replication studies. Estimation uses data from original studies and not replication studies. This allows for `out-of-sample' replication rate predictions. Accurate predictions provide evidence that issues with power outlined in Proposition 3 can adequately explain low observed replication rates. Discrepancies suggest factors not included in the model may also be important.

\subsection{Replication Studies}
I examine two replication studies. \citet{Camerer2016} replicate results from all 18 between subjects laboratory experiments published in \textit{American Economic Review} and \textit{Quarterly Journal of Economics} between 2011 and 2014. \citet{OpenScience2015} replicate results from 100 psychology studies in 2008 from \textit{Psychological Science}, \textit{Journal of Personality and Social Psychology}, and \textit{Journal of Experimental Psychology: Learning, Memory, and Cognition}. Following \citet{Andrews2019}, I consider a sample of 73 psychology studies with test statistics that are well-approximated by $z$-statistics.

In \citet{Camerer2016}, replicators set power rule to detect original effects with at least 90\% power. In \citet{OpenScience2015}, replication teams were instructed to achieve at least 80\% power, and encouraged to obtain higher power if feasible. Reported mean intended power was 92\% in both applications.

\subsection{Estimation}

To calculate the expected replication rate in practice, it is necessary to estimate the latent distribution of studies $\mu_{\Theta,\Sigma}$. To do this, I estimate an augmented version of the empirical model in \citet{Andrews2019}. Specifically, \citet{Andrews2019} develop an empirical model to estimate the marginal distribution of true effects $\Theta^*$, but not the standard errors $\Sigma^*$. Since estimates of the replication rate also require knowledge of the distribution of $\Sigma^*$, I augment the model to estimate the joint distribution of $(\Theta^*,\Sigma^*)$. Estimation is based on the `metastudy approach', which only uses data from original studies. Identification requires that true effects are statistically independent of standard errors, a common assumption in meta-analyses. As in \citet{Andrews2019}, I assume $\Theta^*$ follows a gamma distribution. I additionally assume that $\Sigma^*$ follows a gamma distribution. There are two critical value cutoffs, at 1.64 and 1.96. This gives the following model

$$
|\Theta^*| \sim \Gamma (\kappa_\theta, \lambda_\theta)
$$
$$
\Sigma^* \sim \Gamma (\kappa_\sigma, \lambda_\sigma)
$$
$$
\Theta^* \independent \Sigma^*
$$

\begin{equation}
r(X/\Sigma) \times p(X/\Sigma) \propto     \begin{cases} 
          \beta_{p1} & \text{if } 	\big|\frac{X}{\Sigma}\big| < 1.64 \\
          \beta_{p2} & \text{if } 	1.64 \leq \big|\frac{X}{\Sigma}\big| < 1.96 \\
          1 & \text{if } \big|\frac{X}{\Sigma}\big| \geq 1.96  \\
    \end{cases}
    \label{eq:model}
\end{equation}

Calibrating the model requires that we specify up to unknown parameters the joint probability of being published and chosen for replication. Separate identification of the publication probability function, $p()$, requires that we specify the replication selection function $r()$. 

Replication selection differs across studies. \citet{OpenScience2015} select the last experiment in chosen studies for replication. I treat this selection mechanism as effectively random with respect to the $t$-ratio. This implies that the coefficients ($\beta_{p1}, \beta_{p2})$ measure relative publication probabilities based on statistical significance.

By contrast, \citet{Camerer2016} select the most important significant result, as emphasized by authors, for replication. Of the 18 replication studies, 16 had $p$-values below 0.05 and two had $p$-values slightly above 0.05, but were treated as `positive' results for replication.\footnote{The two studies with $p$-values just above 0.05 had absolute $t$-ratios of 1.93 and 1.81.} I assume replication selection is random with respect to the $t$-ratio for results whose $p$-values are below or only slightly above 0.05. This implies that $\beta_{p2}$ measures the relative probability of being published and chosen for replication for a result whose $p$-value is slightly above 0.05, compared to if it were strictly below 0.05. Only significant (or `almost' significant) results were chosen for replication, so $\beta_{p1}$ equals zero for economics and is not estimated. Further details on replication selection are in Appendix E.

The implied marginal likelihood of $(x, \sigma)$ is

\begin{equation}
    f_{X, \Sigma}(x, \sigma)
    =
    \frac{ p\big(\frac{x}{\sigma} \big) r\big(\frac{x}{\sigma} \big) \int_{\theta} \frac{1}{\sigma} \phi \Big(\frac{x - \theta}{\sigma} \Big) dF_{\Theta}(\theta | \kappa_\theta, \lambda_\theta)
    . 
    g_{\Sigma}(\sigma | \kappa_\sigma, \lambda_\sigma)
    }
    { 
    \int_{\sigma} \Bigg(
    \int_{x'} p\big(\frac{x'}{\sigma}) r\big(\frac{x'}{\sigma} \big)
    \int_{\theta}
    \frac{1}{\sigma} \phi \Big(\frac{x' - \theta}{\sigma} \Big) dx' dF_{\Theta}(\theta | \kappa_\theta, \lambda_\theta)
    \Bigg)
    g_{\Sigma}(\sigma | \kappa_\sigma, \lambda_\sigma) d\sigma
    } 
    \label{eq:econ_likelihood}
\end{equation}

\noindent
where $\phi()$ is the standard normal density, $F_{\Theta}$ is the gamma distribution function for $|\Theta^*|$, and $g_\Sigma$ is the gamma density function for $\Sigma^*$. Assuming independence across studies, the log-likelihood of the data $\{ x_i, \sigma_i \}_i$ is $\ell (\kappa_\theta, \lambda_\theta, \kappa_\sigma, \lambda_\sigma, \beta_p) = \sum_{i} \log{ f_{X, \Sigma}(x_{i}, \sigma_{i}) }$, where $\beta_p$ is a vector of the parameters of the publication probability function. Maximum likelihood estimates are presented in Table \ref{tab:mle_results}.

\begin{table}[H] \centering 
  \footnotesize
  \caption{\textsc{Maximum Likelihood Estimates}}
\begin{tabular}{@{\extracolsep{5pt}} lcccc|cc} 
\\[-1.8ex]\hline 
\hline \\[-1.8ex] 
 & $\kappa_\theta$ & $\lambda_\theta$ & $\kappa_\sigma$ & $\lambda_\sigma$ & $\beta_{p1}$ & $\beta_{p2}$\\ 
\hline \\[-1.8ex] 
Economics experiments  & 1.426 & 0.148  & 2.735 & 0.103 & 0 & 0.038 \\ 
 & (1.298) & (0.073) & (0.538) &  (0.032) & -- & (0.050) \\ 
& & & & & \\
\hline \\[-1.8ex] 
Psychology experiments      & 0.906 & 0.156  &  4.762 & 0.044 & 0.012 & 0.299 \\ 
        & (0.523) & (0.055) & (0.624) &  (0.008) & (0.007) & (0.131) \\ 
\hline
\hline
\end{tabular} 
\caption*{\textit{Notes}: Maximum likelihood estimates for \citet{Camerer2016} and \citet{OpenScience2015}. Robust standard errors are in parenthesis. Joint publication and replication probability coefficients are measured relative to the omitted category of studies significant at 5 percent level. For example, in experimental economics, an estimate of $\beta_{p2}=0.038$ implies that results which are significant at the 5\% level are 26.3 times more likely to be published and chosen for replication than results that are significant at the 10\% level but insignificant at the 5\% level.}
\label{tab:mle_results}
\end{table}

\subsection{The Simulated Replication Rate}
The estimated models in Table \ref{tab:mle_results} are used to generate simulated replication rate predictions, which are then compared to observed replication rates. Replication probabilities depend on power calculations. Following what was actually implemented in \citet{Camerer2016} and \citet{OpenScience2015}, I assume that replicators use the common power rule to set replication standard errors. Specifically, I assume the replicator sets $\sigma_r$ such that the replication probability would equal intended power target $1-\beta^n$ if the true effect were equal to the original published estimate. I set $1-\beta^n = 0.92$ to match the mean intended power reported in both replication studies.\footnote{In practice, intended power for individual replications varied around mean intended power for at least two reasons. First, replication teams were instructed to meet minimum levels of statistical power, and encouraged to obtain higher power if feasible. Second, a number of replications in \citet{OpenScience2015} did not meet this requirement. Appendix F reports simulated replication rates allowing for variation in intended power across studies. Results are largely unchanged.} The procedure is as follows:

\begin{enumerate}
    \item Draw $10^7$ latent (published or unpublished) research questions and standard errors $(\theta^{*sim}, \sigma^{*sim})$ from the estimated joint distribution $\hat{\mu}_{\Theta, \Sigma}$.
    \item Draw estimated effects $x^{*sim}|\theta^{*sim}, \sigma^{*sim} \sim N(\theta^{*sim}, \sigma^{*sim2})$ for each latent study.
    \item Use the estimated selection parameters $(\hat{\beta}_{p1}, \hat{\beta}_{p2})$ to determine the subset of studies that are published and chosen for replication. 
    \item For the subset of replication studies, calculate the replication standard error $\sigma_r^{sim}$ according to the common power rule
    \begin{equation}
        \sigma_r^{sim}(x^{sim}, \beta^n) = \frac{|x^{sim}|}{1.96-\Phi^{-1}(\beta^n)}
    \end{equation}
    \item Draw a replication effect size $x_r^{sim}|\theta^{sim}, \sigma_r^{sim} \sim N(\theta^{sim}, \sigma_r^{sim2})$
\end{enumerate}

Let $\{x_i, \sigma_i, x_{r,i}, \sigma_{r,i} \}_{i=1}^{M_{sig}}$ be the (simulated) set of published, replicated original studies that are significant at the 5\% level, and their corresponding replication results. $M_{sig}$ is the size of the set. The Simulated Replication Rate is equal to

\begin{equation}
    \frac{1}{M_{sig}} \sum_{i=1}^{M_{sig}}    
    \mathbbm{1} \Big(
    |x_{r,i}| \geq 1.96 \sigma_{r,i}, \text{sign}(x_{r,i}) = \text{sign}(x_i)
    \Big)
    \label{eq:mcrr}
\end{equation}

\subsection{Results}

Table \ref{tab:sim_results_econ} presents the results. In experimental economics, the predicted replication rate is 60.1\%, which is very close to the observed rate of 61.1\%. This suggests that issues with common power calculations can explain essentially the entire gap between observed and target replication rates in experimental economics, even in a simple model without treatment effect heterogeneity, researcher manipulation, or measurement error. The second row shows the results for psychology, where the model predicts a replication rate of 53.9\%. This is well below mean intended power of 92\%, but higher than the observed replication rate of 35.6\%. In this case, the model can account for two-thirds of the replication rate gap.\footnote{In both applications, a small number of original results whose $p$-values were slightly above 0.05 were treated as `positive' results and included in the replication rate calculation. For clarity, the theory and the main empirical results maintain a strict 5\% significance threshold for inclusion in the replication rate; however, simulation results are robust to including results with $p$-values just above the 0.05 in the  replication rate.} Results examining alternative rules for setting replication power are presented in Appendix F.

\begin{table}[H] \centering 
\small
  \caption{\textsc{Replication Rate: Observed and Simulated}}
\begin{tabular}{lc|cc} 
\\[-1.8ex]\hline 
\hline \\[-1.8ex] 
                        & Intended power & Observed & Simulated  \\ 
\hline \\[-1.8ex] 
Economics experiments  & 0.92 & 0.611 & 0.601  \\ 
Psychology experiments & 0.92 & 0.356 & 0.539 \\ 
\hline
\hline
\end{tabular} 
\caption*{\textit{Notes}: Economics experiments refer to \citet{Camerer2016} and psychology experiments to \citet{OpenScience2015}. The replication rate is defined as the share of original estimate whose replications have statistically significant findings of the same sign. Figures in the first column report the mean intended power reported in both applications. The second column shows the observed replication rate. The third column reports the simulated replication rate in equation \eqref{eq:mcrr} calculated using parameter estimates Table \ref{tab:mle_results}.}
\label{tab:sim_results_econ}
\end{table} 

The unexplained portion of the gap suggests that issues with common power calculations can account for some but not all of the replication rate gap in psychology. Other factors discussed in the literature and not incorporated in the model may be important, including heterogeneity in true effects, forms of p-hacking not captured in the model, and measurement error. It is also possible that features of the current model are misspecified in psychology. For example, the distribution of true values or standard errors may not be well approximated by a gamma function, true values and standard errors may not be statistically independent, or the publication bias function $p()$ could have another form.

\subsection{Alternative Measures of Publication Bias}
We have shown that the replication rate is unresponsive to the most salient form of publication bias. For journals and policymakers seeking to change current norms, this highlights the need for more informative measures. A range of alternative approaches that better measure the presence and consequences of publication bias have been proposed \citep{Egger1997, Ioannidis2008, Gelman2014, Andrews2019, Stanley2021}. 

In this section, I conduct policy simulations using the estimated model to show how two of these measures -- mean bias and coverage -- change with publication bias. Simulations assume that all results significant at the 5\% level are published, and that results insignificant at the 5\% level are published with probability $\beta_p$. Policymakers' successful efforts to increase the probability of publishing null results leads to an increase in the policy variable, $\beta_p$. Note that while model estimation assumes two cutoffs (1.64 and 1.96), policy simulations are performed assuming policymakers influence publication probabilities at a single cutoff (1.96) for simplicity (i.e. in the policy simulations I set $\beta_p = \beta_{p1} = \beta_{p2}$).

\begin{figure} [H]
	\centering
\includegraphics[width=0.805\textwidth]{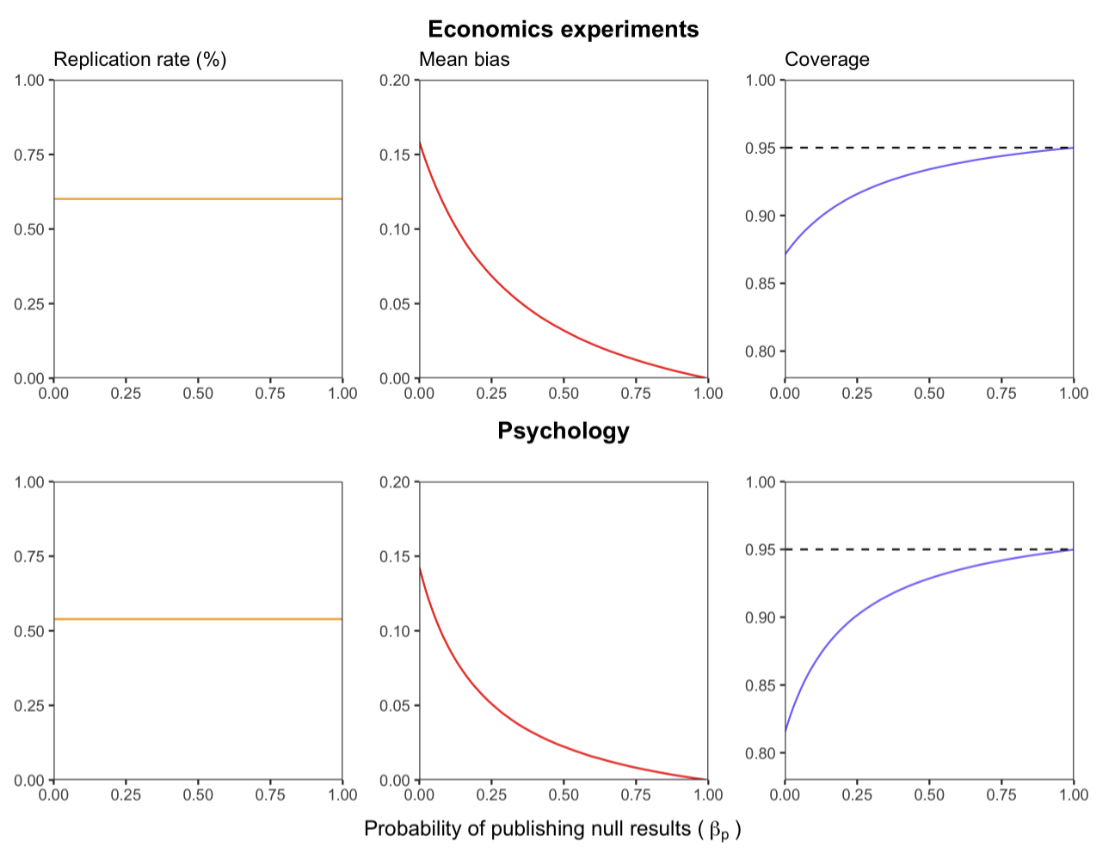} 
	\caption{\textsc{Policy Simulations -- The Replication Rate, Mean Bias, and Coverage}}
	\caption*{\textit{Notes:} The replication rate is defined as the share of replication for which are statistically significant and have the same sign as the original study, mean bias as $\mathbbm{E}(X^* - \Theta^*| D=1)$, and coverage as $\mathbbm{P}(\Theta^* \in (X^* - 1.96 \Sigma^*, X^* + 1.96 \Sigma^*| D=1)$. Results are based on Monte Carlo simulations using model estimates of the latent distribution of studies from Table \ref{tab:mle_results} and setting different levels of publication bias $\beta_p$. The first column reports simulated replication rate predictions based on realized intended power in Table \ref{tab:sim_results_econ}. In the second column, mean bias is in Fisher-transformed correlation coefficient units. In the third column, the horizontal dashed line represents nominal coverage of 0.95. Publication probability $\beta_p$ is measured relative to the omitted category of studies significant at 5 percent level, so $\beta_p=0.1$ implies that results which are significant at the 5 percent are 10 times more likely to be published than insignificant results.}
	\label{fig:bias_and_coverage}
\end{figure}

Figure \ref{fig:bias_and_coverage} shows the results. When few null results are published, mean bias is high and coverage is low. As a benchmark for the mean bias figures, note that the mean effect size in replications is 0.307 in economics and 0.223 in psychology (in Fisher-transformed correlation coefficient units). As fewer null results are censored in the publication process, mean bias approaches zero and actual coverage approaches nominal coverage. The curvature of these graphs is also informative. When publication bias is severe, even small increases in the probability of publishing insignificant results have relatively large impacts on reducing bias and increasing coverage. In line with the theoretical results in this article, the replication rate is unresponsive to changes in the probability of publishing null results. The replication rate is thus a poor measure to evaluate efforts to reduce publication bias; alternative measures, such as mean bias or coverage, are therefore needed to gauge the effectiveness of such efforts. 

\section{Conclusion}
The prominence of the replication rate stems in part from its apparent transparency and ease of interpretation. However, in reality, it poses substantial difficulties for inference and interpretation. This article shows that the replication rate provides limited information about publication bias, and induces regression to the mean in replications. This and other issues with common power calculations lead inevitably to low replication rates. Accounting for these issues in an empirical model explains the entirety of the replication gap in experimental economics. In light of this, a reassessment of the empirical content of the replication rate is necessary. For those interested in measuring publication bias, alternative statistics more responsive to the prevalence of published null results should be used over the replication rate.

\bigskip
\bibliographystyle{aer}
{\small
\bibliography{References}

@article{Card1995,
author = {Card, David and Krueger, Alan B.},
journal = {American Economic Review: Papers and Proceedings},
doi = {},
volume = {85},
number = {2},
pages = {238-243},
primaryClass = {},
title = {Time-Series Minimum-Wage Studies: A Meta-analysis},
year = {1995}
}

@article{Ioannidis2005,
author = {Ioannidis, John P.A.},
journal = {PLoS Med},
doi = {https://doi.org/10.1371/journal.pmed.0020124},
volume = {2},
number = {8},
pages = {},
primaryClass = {},
title = {{Why Most Published Research Findings Are False}},
year = {2005}
}

@article{Andrews2019,
author = {Andrews, Isaiah and Kasy, Maximilian},
journal = {American Economic Review},
doi = {},
volume = {109},
number = {8},
pages = {2766-2794},
primaryClass = {},
title = {{Identification of and Correction for Publication Bias}},
year = {2019}
}

@article{Ioannidis2008,
author = {Ioannidis, John P.A.},
journal = {Epidemiology},
doi = {doi: 10.1097/EDE.0b013e31818131e7.},
volume = {19},
number = {5},
pages = {640–648},
title = {{Why Most Discovered True Associations Are Inflated}},
year = {2008}
}

@article{Gelman2014,
author = {Gelman, Andrew and Carlin, John},
journal = {Perspectives on Psychological Science},
volume = {9},
number = {6},
pages = {641-651},
title = {Beyond Power Calculations: Assessing Type S (Sign) and Type M (Magnitude) Errors},
year = {2014}
}

@article{Gelman2018,
author = {Gelman, Andrew},
journal = {Personality and Social Psychology Bulletin},
doi = {},
volume = {44},
number = {1},
pages = {16-23},
title = {The Failure of Null Hypothesis Significance Testing When Studying Incremental Changes, and What to Do About It
},
year = {2018}
}

@article{Camerer2016,
author = {Camerer, Colin F. and Dreber, Anna and others},
journal = {Science},
number = {6280},
pages = {1433--1437},
title = {{Evaluating replicability of laboratory experiments in economics}},
volume = {351},
year = {2016}
}

@article{Camerer2018,
author = {Camerer, Colin F. and Dreber, Anna and others},
journal = {Nature},
pages = {637–644},
title = {{Evaluating the replicability of social science experiments in Nature and Science between 2010 and 2015}},
volume = {2},
year = {2018}
}

@article{OpenScience2015,
author = {{Open Science Collaboration}},
doi = {10.1126/science.aac4716},
issn = {10959203},
journal = {Science},
number = {6251},
pmid = {26315443},
title = {{Estimating the reproducibility of psychological science}},
volume = {349},
year = {2015}
}

@article{Kasy2022,
author = {Frankel, Alexander and Kasy, Maximilian},
journal = {American Economic Journal: Microeconomics},
doi = {},
volume = {14},
number = {1},
pages = {1-38},
title = {{Which Findings Should Be Published?}},
year = {2022}
}

@article{Kasy2021,
author = {Kasy, Maximilian},
journal = {Journal of Economic Perspectives},
doi = {},
volume = {35},
number = {3},
pages = {175-192},
title = {{Of Forking Paths and Tied Hands: Selective Publication of Findings, and What Economists Should Do about It}},
year = {2021}
}

@article{Galton1886,
author = {Galton, Francis},
journal = {The Journal of the Anthropological Institute of Great Britain and Ireland},
doi = {},
volume = {15},
number = {},
pages = {246-263},
title = {{Regression Towards Mediocrity in Hereditary Stature}},
year = {1886}
}

@article{Abadie2020,
author = {Abadie, Alberto},
journal = {American Economic Review: Insights},
doi = {},
volume = {2},
number = {2},
pages = {193-208},
title = {{Statistical Nonsignificance in Empirical Economics}},
year = {2020}
}

@article{Anderson2017,
author = {Anderson, Smantha F. and Maxwell, Scott E.},
journal = {Multivariate Behavioral Research},
doi = {},
volume = {52},
number = {3},
pages = {305-324},
title = {{Addressing the ``Replication Crisis'': Using Original Studies to Design Replication Studies with Appropriate Statistical Power
}},
year = {2017}
}

@article{Stanley2018,
author = {Stanley, T.D. and Carter, Evan C. and Doucouliagos, Hristos},
journal = {Psychological Bulletin},
doi = {},
volume = {144},
number = {12},
pages = {1325-1346},
title = {{What Meta-Analyses Reveal About the Replicability of Psychological Research
}},
year = {2018}
}

@article{Maxwell2015,
author = {Maxwell, Scott E. and Lau, Michael Y. and Howard, George S.},
journal = {American Psychologist},
doi = {},
volume = {70},
number = {6},
pages = {487-498},
title = {{Is psychology suffering from a replication crisis? What does “failure to replicate” really mean?
}},
year = {2015}
}

@article{Klein2014,
author = {Klein, Richard A. and Ratliff, Kate A. and Vianello, Michelangelo and others},
doi = {10.1027/1864-9335/a000178},
journal = {
Social Psychology},
volume = {45},
number = {3},
pages = {142-152},
title = {{Investigating Variation in Replicability: A ``Many Labs'' Replication Project}},
year = {2014}
}

@article{Klein2018,
author = {Klein, Richard A. and Vianello, Michelangelo and Hasselman, Fred and others},
doi = {https://doi.org/10.1177/2515245918810225},
journal = {Advances in Methods and Practices in Psychological Science},
volume = {1},
number = {4},
pages = {443-490},
title = {{Many Labs 2: Investigating Variation in Replicability Across Samples and Settings}},
year = {2018}
}

@article{Amrhein2019,
author = {Amrhein, Valentin and Trafimow, David and Greenland, Sander},
journal = {The American Statistician},
doi = {},
volume = {73},
number = {1},
pages = {262-270},
primaryClass = {},
title = {Inferential Statistics as Descriptive Statistics: There Is No Replication Crisis if We Don’t Expect Replication},
year = {2019}
}

@article{AmrheinNature2019,
author = {Amrhein, Valentin and Greenland, Sander and McShane, Blake},
journal = {Nature},
doi = {},
volume = {567},
number = {},
pages = {305-307},
primaryClass = {},
title = {Retire Statistical Significance},
year = {2019}
}

@article{McShane2019,
author = {McShane, Blakeley B. and Gal, David and Gelman, Andrew and Robert, Christian and Tackett, Jennifer L.},
journal = {The American Statistician},
doi = {},
volume = {73},
number = {1},
pages = {235-245},
primaryClass = {},
title = {Abandon Statistical Significance},
year = {2019}
}

@article{Shrout2018,
author = {Shrout, Patrick E. and Rodgers, Jospeh L.},
journal = {Annual Review of Psychology},
doi = {},
volume = {69},
number = {},
pages = {487–510},
primaryClass = {},
title = {Psychology, Science, and Knowledge Construction: Broadening Perspectives from the Replication Crisis},
year = {2018}
}

@article{Miguel2018,
author = {Miguel, Edward and Christensen, Garret},
journal = {Journal of Economic Literature},
doi = {},
volume = {56},
number = {3},
pages = {920–980},
primaryClass = {},
title = {Transparency, Reproducibility, and the Credibility of Economics Research},
year = {2018}
}

@article{Loken2017,
author = {Loken, Eric and Gelman, Andrew},
journal = {Science},
doi = {},
volume = {355},
number = {6325},
pages = {584-585},
primaryClass = {},
title = {Measurement error and
the replication crisis},
year = {2017}
}

@article{Simonsohn2014,
author = {Simonsohn, Uri and Nelson, Leif D. and Simmons, Joseph P.},
journal = {Journal of Experimental Psychology: General},
doi = {},
volume = {143},
number = {2},
pages = {534–547},
primaryClass = {},
title = {P-Curve: A Key to the File-Drawer},
year = {2014}
}

@article{Brodeur2016,
author = {Brodeur, Abel and Lé, Mathias and Sangnier, Marc and Zylberberg, Yanos},
journal = {American Economic Journal: Applied Economics},
doi = {},
volume = {8},
number = {1},
pages = {1-32},
primaryClass = {},
title = {Star Wars: The Empirics Strike Back},
year = {2016}
}

@article{Cesario2014,
author = {Cesario, Joseph},
journal = {Perspectives on Psychological Science},
doi = {},
volume = {9},
number = {1},
pages = {40-48},
primaryClass = {},
title = {Priming, Replication, and the Hardest Science},
year = {2014}
}

@article{Simons2014,
author = {Simons, Daniel J.},
journal = {Perspectives on Psychological Science},
doi = {},
volume = {9},
number = {1},
pages = {76–80},
primaryClass = {},
title = {The Value of Direct Replication},
year = {2014}
}

@article{Higgins2002,
author = {Higgins, Julian P.T. and Thompson, Simon G.},
journal = {Statistics in Medicine},
doi = {},
volume = {21},
number = {11},
pages = {1539-1558},
primaryClass = {},
title = {Quantifying heterogeneity in a meta-analysis},
year = {2002}
}

@article{Nature2020,
author = {Editorial},
journal = {Nature},
doi = {},
volume = {578},
number = {},
pages = {489-490},
primaryClass = {},
title = {In praise of replication studies and null results},
year = {2020}
}

@article{Chambers2013,
author = {Chambers, Christopher D.},
journal = {Cortex},
doi = {},
volume = {49},
number = {3},
pages = {609–610},
primaryClass = {},
title = {Registered Reports: A new publishing initiative at Cortex},
year = {2013}
}

@article{Foster2019,
author = {Foster, Andrew and Karlan, Dean and Miguel, Edward and Bogdanoski, Aleksandar},
journal = {World Bank Development Impact Blog},
doi = {},
volume = {},
number = {},
pages = {},
primaryClass = {},
title = {Pre-results Review at the Journal of
Development Economics: Lessons
Learned So Far},
year = {2019}
}

@article{Mervis2014,
author = {Mervis, Jeffrey},
journal = {Science},
doi = {},
volume = {345},
number = {},
pages = {992},
primaryClass = {},
title = {Why null results rarely see the light of day},
year = {2014}
}

@article{Franco2014,
author = {Franco, Annie and Neil Malhotra and Simonovits, Gabor},
journal = {Science},
doi = {},
volume = {345},
number = {6203},
pages = {1502-1505},
primaryClass = {},
title = {Publication bias in the social sciences: Unlocking the file drawer},
year = {2014}
}

@article{Landis2014,
author = {Landis, Ronald S. and James, Lawrence R. and Lance, Charles E. and Pierce, Charles A. and Rogelberg, Steven G.},
journal = {Journal of Business and Psychology },
doi = {},
volume = {29},
number = {},
pages = {163–167},
primaryClass = {},
title = {When is Nothing Something? Editorial for the Null Results Special Issue of Journal of Business and Psychology},
year = {2014}
}

@article{Barnett2004,
author = {Barnett, Adrian G. and Van Der Pols, Jolieke C. and Dobson, Annette J.},
journal = {Journal of Business and Psychology},
doi = {},
volume = {34},
number = {1},
pages = {215–220},
primaryClass = {},
title = {Regression to the mean: what it is and how to deal with it},
year = {2004}
}

@book{Kahneman2011,
  author = {Kahneman, Daniel},
  year = {2011},
  title = {Thinking, Fast and Slow},
  publisher = {Farrar, Straus and Giroux}
}

@article{Stanley2017,
author = {Ioannidis, John P. A. and Stanley, T. D.  and Doucouliagos, Hristos},
journal = {The Economic Journal},
doi = {},
volume = {127},
number = {605},
pages = {236–265},
primaryClass = {},
title = {The Power of Bias in Economics Research},
year = {2017}
}

@article{Tackett2019,
author = {Tackett, Jennifer L. and Brandes, Cassandra M. and King, Kevin M. and Markon, Kristian E.},
journal = {Annual Review of Clinical Psychology},
doi = {},
volume = {15},
number = {},
pages = {579-604},
primaryClass = {},
title = {Psychology's Replication Crisis and Clinical Psychological Science},
year = {2019}
}

@article{Button2013,
author = {Button, Katherine S. and Ioannidis, John and Mokrysz, Claire and others},
journal = {Nature reviews neuroscience},
doi = {},
volume = {14},
number = {5},
pages = {365-376},
primaryClass = {},
title = {Power failure: why small sample size undermines the reliability of neuroscience},
year = {2013}
}

@book{Rothstein2006,
  author = {Rothstein, Hannah R. and Sutton, Alexander J. and Borenstein, Michael},
  year = {2006},
  title = {Publication bias in meta-analysis: Prevention, assessment and adjustments},
  publisher = {John Wiley \& Sons}
}

@article{Benjamin2018,
author = {Benjamin, Daniel J. and Berger, James O. and Johnson, Valen E.},
journal = {Nature Human Behaviour},
doi = {},
volume = {2},
number = {},
pages = {6-10},
primaryClass = {},
title = {Redefine statistical significance},
year = {2018}
}

@book{Miguel2019,
  author = {Christensen, Garret and Freese, Jeremy and Miguel, Edward},
  year = {2019},
  title = {Transparent and Reproducible Social Science Research},
  publisher = {University of California Press}
}

@article{Fisher1915,
author = {Fisher, Ronald A.},
journal = {Biometrika},
doi = {},
volume = {10},
number = {4},
pages = {507-521},
primaryClass = {},
title = {Frequency Distribution of the Values of the Correlation Coefficient in Samples from an Indefinitely Large Population},
year = {1915}
}

@article{Gorroochurn2007,
author = {Gorroochurn, Prakash and Hidge, Susan E. and others},
journal = {Genet Med},
doi = {},
volume = {9},
number = {6},
pages = {325–331},
primaryClass = {},
title = {Non-replication of association studies: “pseudo-failures” to replicate?},
year = {2007}
}

@article{Hotelling1933,
author = {Hotelling, Harold},
journal = {Journal of the American Statistical Association},
doi = {},
volume = {28},
number = {184},
pages = {463-465},
primaryClass = {},
title = {Review: The Triumph of Mediocrity in Business, By Horace Secrist},
year = {1933}
}

@article{Bryan2019,
author = {Bryan, Christopher J. and Yeager, David S. and O’Brien, Joseph M.},
journal = {Proceedings of the National Academy of Sciences of the United States of America},
doi = {},
volume = {116},
number = {51},
pages = {25535-25545},
primaryClass = {},
title = {Replicator degrees of freedom allow publication of misleading failures to replicate},
year = {2019}
}

@article{Stanley2021,
author = {Stanley, T.D. and Doucouliagos, Hristos and Ioannidis, John P.A. and Carter, Evan C.},
journal = {Research Synthesis Methods},
doi = {},
volume = {12},
number = {6},
pages = {776-795},
title = {{Detecting publication selection bias through excess statistical significance
}},
year = {2021}
}

@article{Egger1997,
author = {Egger, Matthias and Smith, George D. and Schneider, Martin and Minder, Christoph},
journal = {Research Synthesis Methods},
doi = {},
volume = {315},
number = {},
pages = {629-634},
title = {{Bias in meta-analysis detected by a simple, graphical test
}},
year = {1997}
}
}

\newpage
\section*{Appendix}
\section*{A. Properties of the Replication Probability Function}

This Appendix derives properties of the replication probability function (Definition 1). The first `property' simply provides a convenient, compact notation. The remaining properties consider the replication probability function under the common power rule to detect original effect sizes with $1-\beta^n$ intended power (Definition 3). Recall that the replication probability for original study $(x, \sigma, \theta)$ is equal to

\begin{equation}
    RP\big( x, \theta, \sigma_r(x, \sigma, \beta^n) \big) =     \mathbbm{P} \Bigg( \frac{| {X}_r | }{\sigma_r(x, \beta^n)} \geq 1.96 ,  \text{sign}(X_r) = \text{sign}(x) \Bigg)
    \label{eq:app_a_cond_rp}
\end{equation}

\begin{figure} [H]
	\centering
\includegraphics[width=0.8\textwidth]{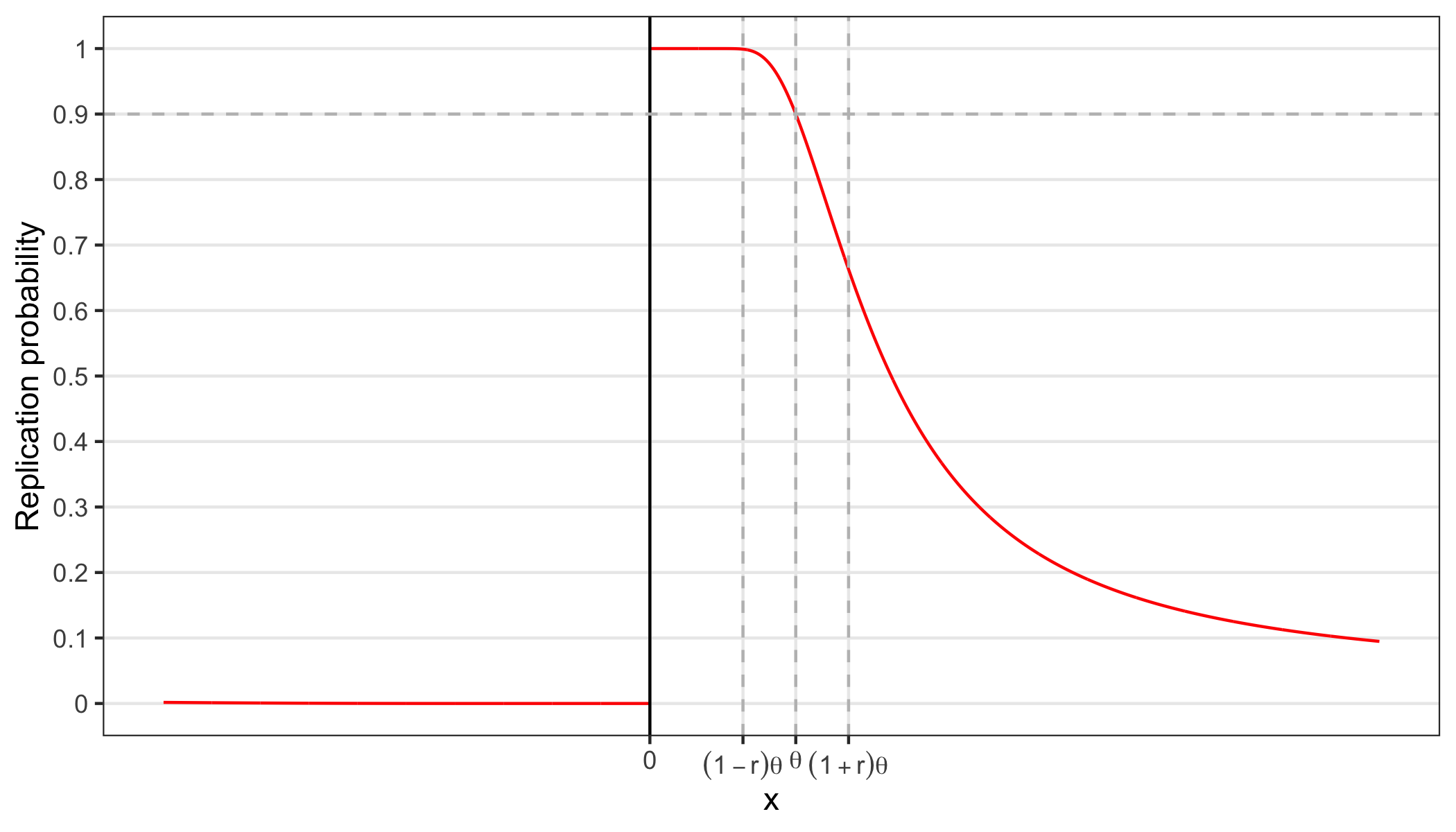} 
	\caption{Example of the replication probability function under the common power rule with intended power $(1-\beta^n)=0.9$. The two vertical lines around $\theta$ marks the open interval over which the replication probability function is strictly concave, where $r^*$ is given by equation \eqref{eq:app_a_quad_solution}. }
	\label{fig:app_a_rep_prob}
\end{figure}

\noindent
To provide intuition of the properties, Figure \ref{fig:app_a_rep_prob} provides an illustration of the replication probability function for different values of $x$ under the common power rule for $1-\beta^n = 0.9$ and a fixed value of $\theta$. \\

\noindent
\textbf{Lemma 2} (Properties of the replication probability function). \textit{The replication probability function satisfies the following properties:}
\begin{enumerate}
    \item \textit{For any replication standard error $\sigma_r(x, \sigma, \beta^n)$, the replication probability for an original study $(x, \sigma, \theta)$ can be written compactly as}
        \begin{equation}
            RP\big( x, \theta, \sigma_r(x, \sigma, \beta^n) \big) =  1 -  \Phi \bigg( 1.96 - \text{sign}(x) \frac{\theta}{\sigma_r(x, \sigma, \beta^n)} \bigg)
            \label{eq:app_a_1}
        \end{equation}
\end{enumerate}

\noindent
The remaining properties assume the replication standard error $\sigma_r(x, \beta^n)$ is set using the common power rule in Definition 3 with intended power $1-\beta^n$: \\
\begin{enumerate}
    \setcounter{enumi}{1}
    \item \textit{If $1-\beta^n>0.025$, then $RP\big( x, \theta, \sigma_r(x, \beta^n) \big)$ is strictly decreasing in $x$ over $(-\infty, 0)$ and $(0, \infty)$. }
    
    \item \textit{If $(1-\beta^n)>0.6628$, then $RP\big( x, \theta, \sigma_r(x, \beta^n) \big)$ is strictly concave with respect to $x$ over the open interval $( \max{\{0, [1-r^*(\beta^n)]\theta}\}, [1+r^*(\beta^n)]\theta)$, where}
    \begin{equation}
        r^*(\beta^n) = - \big(2 + 1.96.h(\beta^n) \big) + 
        \sqrt{ \frac{\big(2 + 1.96.h(\beta^n)\big)^2 - 4 \times (1+1.96.h(\beta^n) - h(\beta^n)^2\big)}{2} } >0
        \label{eq:app_a_quad_solution}
    \end{equation}
    \noindent
    with $h(\beta^n) = \big(1.96 - \Phi^{-1}(\beta^n)\big)$.
    \item \textit{The limits of the replication probability function with respect to $x$ are}
    \begin{equation}
        \lim_{x\to\infty} RP\big( x, \theta, \sigma_r(x, \beta^n) \big)=  0.025 \text{ and }  \lim_{x\to-\infty} RP\big( x, \theta, \sigma_r(x, \beta^n) \big)=  0.025
    \end{equation}
    \begin{equation}
        \lim_{x \uparrow 0} RP\big( x, \theta, \sigma_r(x, \beta^n) \big) =  0 \text{ and }
        \lim_{x \downarrow 0} RP\big( x, \theta, \sigma_r(x, \beta^n) \big) =  1
    \end{equation}
\end{enumerate}
    
\subsubsection*{Proof of 1.}

The probability in equation \eqref{eq:app_a_cond_rp} equals $\big[ \mathbbm{1}(x/\sigma \geq 1.96) \times  \big( 1 -  \Phi \big( 1.96 - \frac{\theta}{\sigma_r} \big)\big] + \big[ \mathbbm{1}(x/\sigma \leq -1.96) \times  \Phi \big( -1.96 - \frac{\theta}{\sigma_r} \big)\big]$. This captures the two requirements for `successful' replication: the replication estimate must attain statistical significance and have the same sign as the original estimate. Equation \eqref{eq:app_a_1} is obtained using the symmetry of the normal distribution, which implies that $\Phi(t) =  1 - \Phi(-t)$ for any $t$. $\square$

\subsubsection*{Proof of 2.} The first derivative of the replication probability function with the common power rule is

\begin{equation}
\frac{\partial RP\big( x, \theta, \sigma_r(x, \beta^n)\big) }{\partial x} = 
    \begin{cases} 
      - \frac{\theta}{x^2}
\big( 1.96 - \Phi^{-1}(\beta^{n}) \big)
\times
\phi \Big( 1.96  - \frac{\theta}{x} \big( 1.96 - \Phi^{-1}(\beta^{n}) \big) \Big), & x > 0 \\
      - \frac{\theta}{x^2}
\big( 1.96 - \Phi^{-1}(\beta^{n}) \big)
\times
\phi \Big( -1.96  - \frac{\theta}{|x|} \big( 1.96 - \Phi^{-1}(\beta^{n}) \big) \Big), & x <0 
   \end{cases}
\end{equation}
\noindent
These are strictly negative whenever $\big( 1.96 - \Phi^{-1}(\beta^{n}) \big) > 0 \iff (1-\beta^n)>0.025$. $\square$

\subsubsection*{Proof of 3.} First, note that for $x>0$, the second derivative of the replication probability function with the common power rule is

\begin{equation}
    \frac{\partial^2 RP\big( x, \theta, \sigma_r(x, \beta^n) \big)}{\partial x^2} = 
    \bigg( \frac{h(\beta^n) \theta}{x^3} \bigg) 
    \phi\bigg( 1.96 - \frac{h(\beta^n) \theta}{x} \bigg)
    \Bigg[
    1 + \bigg( \frac{h(\beta^n) \theta}{x} \bigg)  \bigg(  1.96 - \frac{h(\beta^n) \theta}{x} \bigg)
    \Bigg]
    \label{eq:app_a_second_derivative}
\end{equation}

Let $x = (1+r)\theta$. Substituting this into the previous equation and simplifying shows that equation \eqref{eq:app_a_second_derivative} is strictly negative when the following inequality is satisfied

\begin{equation}
    r^2 + \big( 2 + 1.96 h(\beta^n) \big). r + \big(1 + 1.96 h(\beta^n) - h(\beta^n)^2 \big) < 0
    \label{eq:app_a_quad}
\end{equation}

The solution to the quadratic equation has a unique positive solution $r^*(\beta^n)$ whenever $(1-\beta^n)>0.6628$. To see this, note that there exists a unique positive solution when $\big(1 + 1.96 h(\beta^n) - h(\beta^n)^2 \big) < 0$. This quadratic equation in $h(\beta^n)$ must have a unique positive and negative solution in turn, since the parabola opens downwards and equals 1 when $h(\beta^n)=0$. The positive root can be obtained from the quadratic formula, which gives $2.38014$. Since the quadratic function opens downward, this implies that for any $h(\beta^n) > 2.38014$, we have $\big(1 + 1.96 h(\beta^n) - h(\beta^n)^2 \big) < 0$. Thus, a unique positive solution to equation \eqref{eq:app_a_quad} exists whenever this condition is satisfied. In particular, a unique positive solution exists whenever

$$
h(\beta^n) = 1.96 - \Phi^{-1}(\beta^n) >2.38014 
$$
$$
\iff \Phi (1.96 - 2.38014) > \beta^n
$$
\begin{equation}
\iff (1-\beta^n) > 0.6628
\end{equation}

The unique positive solution for equation \eqref{eq:app_a_quad} can again be obtained by the quadratic formula, which gives equation \eqref{eq:app_a_quad_solution}. Note that for any $r>0$ where the inequality for concavity in equation \eqref{eq:app_a_quad} is satisfied, the same must also be true of $-r$, since it makes the left-hand-side strictly smaller. This implies that the  replication probability function is strictly concave (since its second derivative is strict negative) over $(  \max{ \{0, [1-r^*(\beta^n)]\theta\}}, [1+r^*(\beta^n)]\theta)$, where the maximum is taken because the replication probability function is discontinuous at 0. This follows because of the properties of the quadratic function. Specifically, suppose $f(x)$ is a parabola that opens upward and intersects the y-axis at a negative value. Then for any two points $(a,b)$ with $a<b$ and $f(a), f(b)<0$, it must be that $f(c)<0$ for any $c \in (a,b)$. $\square$

\subsubsection*{Proof of 4.}  Substituting the common power rule into the replication probability function gives

\begin{equation}
    RP\big( x, \theta, \sigma_r(x, \beta^n)  \big)
    = 1 -  \Phi \bigg( 1.96 - \frac{\theta}{x} \big(1.96 - \Phi^{-1}(\beta^n) \big) \bigg)
\end{equation}

The values of the limits can be seen immediately from this expression. $\square$


\section*{B. Proofs of Propositions}

\noindent
\textbf{Proof of Proposition 2:} We have $\mathbbm{E} \big(X_r \big| \Theta = \theta \big)=\theta$ by assumption. Next, note that

$$
    \mathbbm{E}_{X^*| \Theta^* , S_X^*, D} \Big( X^* | \Theta^* = \theta, |X^*/\Sigma^* | \geq 1.96, D=1 \Big) 
    =
    \mathbbm{E}_{X| \Theta , S_X} \Big( X | \Theta = \theta, |X/\Sigma | \geq 1.96 \Big) 
$$
\begin{equation}
    =
    \mathbbm{E}_{\Sigma | \Theta, S_X} \Bigg( 
    \mathbbm{E}_{X| \Theta, \Sigma, S_X} \Big( X | \Theta = \theta, \Sigma = \sigma, |X/\sigma | \geq 1.96 \Big) 
    \Bigg)
    \label{eq:prop2_1}
\end{equation}

\noindent
where the last line uses the Law of Iterated Expectations. We will prove $\mathbbm{E}_{X| \Theta, \Sigma, S_X^*} \big( X | \Theta = \theta, \Sigma = \sigma, |X/\sigma | \geq 1.96  \big) > \theta$, which implies that equation \eqref{eq:prop2_1} is also greater than $\theta$. Recall that $X|\theta, \sigma$ is the effect size of published studies and follows a truncated normal distribution:

\begin{equation}
\frac{p\big(\frac{x}{\sigma} \big)\frac{1}{\sigma} \phi\big(\frac{x-\theta}{\sigma} \big) \mathbbm{1}\big(|\frac{x}{\sigma} |\geq 1.96\big)} 
{
\int p\big(\frac{x'}{\sigma} \big)\frac{1}{\sigma} \phi\big(\frac{x'-\theta}{\sigma} \big) \mathbbm{1}\big(|\frac{x}{\sigma} |\geq 1.96\big)dx'} 
\end{equation}

Define $X = \theta + \sigma Z$. Then the density for the transformed random variable $Z$ is 

\begin{equation}
    \frac{p\big(z+\frac{\theta}{\sigma} \big) \phi\big(z \big) \mathbbm{1}\big(|z+\frac{\theta}{\sigma}  |\geq 1.96\big)}
    {
    \int p\big(z'+\frac{\theta}{\sigma} \big) \phi\big(z' \big)\mathbbm{1}\big(|z+\frac{\theta}{\sigma}  |\geq 1.96\big)dz'
    }
\end{equation}

For notational convenience, define the following normalization constants:

\begin{equation}
\bar{\eta} = 
\mathbbm{P}(X \leq -1.96\sigma) + \mathbbm{P}(X\geq1.96\sigma) 
= 
\mathbbm{P}\bigg(Z \leq -1.96 - \frac{\theta}{\sigma} \bigg) + \mathbbm{P}\bigg(Z \geq 1.96 - \frac{\theta}{\sigma} \bigg)
\end{equation}

\begin{equation}
\eta_1 = \mathbbm{P}(X \leq -1.96\sigma) 
= \mathbbm{P}\bigg(Z \leq -1.96 - \frac{\theta}{\sigma} \bigg)
\end{equation}

\begin{equation}
\eta_2 = \mathbbm{P}(X \geq 2\theta + 1.96\sigma ) 
= \mathbbm{P}\bigg(Z \geq \frac{\theta}{\sigma} + 1.96 \bigg)
\end{equation}

\begin{equation}
\eta_3 = \mathbbm{P}(1.96 \sigma \leq X \leq 2\theta - 1.96\sigma ) 
= \mathbbm{P}\bigg(1.96 - \frac{\theta}{\sigma} \leq Z \leq \frac{\theta}{\sigma} - 1.96  \bigg)
\end{equation}

\subsubsection*{Case 1.} Consider two cases. First, suppose $\theta \in (0, 1.96\sigma)$. Conditional on $(\theta, \sigma$) (where we suppress the conditional notation on $(\theta, \sigma)$ for clarity), the expected value of a published estimate conditional of statistical significance is

$$
\mathbbm{E}(X | 1.96\sigma \leq |X| )
=
\frac{1}{\bar{\eta}} \Bigg(
\eta_1 \mathbbm{E}(X | X \leq -1.96 \sigma)
+ \eta_2 \mathbbm{E}(X | X \geq 2\theta + 1.96\sigma )
$$
\begin{equation}
+ \big(\bar{\eta}- \eta_1 - \eta_2 \big) \mathbbm{E}(X | 1.96\sigma \leq X \leq 2\theta + 1.96\sigma )
\Bigg)
\end{equation}

First note that $\mathbbm{E}(X | 1.96\sigma \leq X \leq 2\theta + 1.96\sigma )>\theta$ since we assume that $\theta \in (0, 1.96\sigma)$ and $p_{sig}()>0$. If $
\eta_1 \mathbbm{E}(X | X \leq -1.96 \sigma)
+ \eta_2 \mathbbm{E}(X | X \geq 2\theta + 1.96\sigma ) \geq 
\big(\eta_1 + \eta_2 \big) \theta$, it follows that $\mathbbm{E}(X | 1.96\sigma \leq |X|) > \theta$, which is what we want to show. Consider the first expectation in this expression: 

\begin{equation}
\mathbbm{E}(X | X \leq -1.96 \sigma) = \mathbbm{E}\Big(\theta + \sigma Z | Z \leq -1.96 -\frac{\theta}{\sigma} \Big)
=
\theta + \sigma \mathbbm{E}\Big( Z | Z \leq -1.96 -\frac{\theta}{\sigma} \Big)
\label{eq:prop2_2}
\end{equation}

Evaluating the expectation in the right-hand-side of equation \eqref{eq:prop2_2} gives

$$
\mathbbm{E}\Big( Z | Z \leq -1.96 -\frac{\theta}{\sigma} \Big)
=
\frac{1}{\eta_1} \int_{-\infty}^{-1.96 - \frac{\theta}{\sigma}} z p_{sig} \bigg(z + \frac{\theta}{\sigma} \bigg) \phi(z) dz
=
- \frac{1}{\eta_1}\int_{-\infty}^{-1.96 - \frac{\theta}{\sigma}}  p_{sig} \bigg(z + \frac{\theta}{\sigma} \bigg) \phi'(z) dz
$$
$$
=
- \frac{1}{\eta_1} \Bigg[
p_{sig} (-1.96) \phi \bigg( -1.96-\frac{\theta}{\sigma} \bigg)
-
p_{sig}(-\infty) \phi(-\infty)
-
\int_{-\infty}^{-1.96 - \frac{\theta}{\sigma}}  p_{sig} '\bigg(z + \frac{\theta}{\sigma} \bigg) \phi(z) dz
\Bigg]
$$
\begin{equation}
    =
    -\frac{1}{\eta_1} p_{sig}(-1.96) \phi \bigg( -1.96-\frac{\theta}{\sigma} \bigg)
    +
    \frac{1}{\eta_1} \int_{-\infty}^{-1.96 - \frac{\theta}{\sigma}}  p_{sig}'\bigg(z + \frac{\theta}{\sigma} \bigg) \phi(z) dz
    \label{eq:prop2_int}
\end{equation}

\noindent
where the second equality uses $\phi'(z) = -z \phi(z)$; the third equality uses integration by parts; and the final equality follows because $p_{sig}(-\infty) \phi(-\infty)=0$ since we assume $p_{sig}()$ is bounded. Substituting this into equation \eqref{eq:prop2_2} gives

\begin{equation}
    \mathbbm{E}(X | X \leq -1.96 \sigma) = 
    \theta 
    - \frac{\sigma}{\eta_1} p_{sig}(-1.96) \phi \bigg( -1.96-\frac{\theta}{\sigma} \bigg)
    + \frac{\sigma}{\eta_1} \int_{-\infty}^{-1.96 - \frac{\theta}{\sigma}}  p_{sig}'\bigg(z + \frac{\theta}{\sigma} \bigg) \phi(z) dz
    \label{eq:prop2_part1}
\end{equation}

Next, note that 

\begin{equation}
\mathbbm{E}(X | X \geq 2\theta + 1.96\sigma ) =
\theta 
+
\sigma \mathbbm{E}\Big( Z | Z \leq \frac{\theta}{\sigma} + 1.96 \Big)
\label{eq:prop2_ref}
\end{equation}

\noindent
where

\begin{equation}
\mathbbm{E}\Big( Z | Z \leq \frac{\theta}{\sigma} + 1.96 \Big)
=
\frac{1}{\eta_2}
\int_{1.96 + \frac{\theta}{\sigma} }^{\infty}  z p_{sig}\bigg(z + \frac{\theta}{\sigma} \bigg) \phi(z) dz
\geq
\frac{1}{\eta_2}
\int_{1.96 + \frac{\theta}{\sigma} }^{\infty}  z p_{sig}\bigg(z - \frac{\theta}{\sigma} \bigg) \phi(z) dz
\end{equation}

\noindent
since $p_{sig}(z + \theta/\sigma) \geq p_{sig}(z - \theta/\sigma)$ for all $z \in (1.96 + \theta/\sigma, \infty)$ because $p_{sig}(t)$ is weakly increasing over $t>1.96$. For the right-hand-side of this equation, we can apply similar arguments used to derive equation \eqref{eq:prop2_int}. Substituting the result into equation \eqref{eq:prop2_ref} gives 

\begin{equation}
\mathbbm{E}(X | X \geq 2\theta + 1.96\sigma )
\geq
\theta 
+ \frac{\sigma}{\eta_2} p_{sig}(1.96 ) \phi \bigg( 1.96+\frac{\theta}{\sigma} \bigg)
+ \frac{\sigma}{\eta_2} \int_{1.96 + \frac{\theta}{\sigma} }^{\infty}  p_{sig}'\bigg(z - \frac{\theta}{\sigma} \bigg) \phi(z) dz
 \label{eq:prop2_part2}
\end{equation}

Equations \eqref{eq:prop2_part1} and \eqref{eq:prop2_part2} imply

$$
\eta_1 \mathbbm{E}(X | X \leq -1.96 \sigma)
+ \eta_2 \mathbbm{E}(X | X \geq 2\theta + 1.96\sigma )
$$
$$
\geq
(\eta_1 + \eta_2) \theta
+
\sigma \Bigg[
p_{sig}(1.96 ) \phi \bigg( 1.96+\frac{\theta}{\sigma} \bigg) -
p_{sig}(-1.96) \phi \bigg( -1.96-\frac{\theta}{\sigma} \bigg)
\Bigg]
$$
\begin{equation}
+
\sigma \Bigg[
\int_{-\infty}^{-1.96 - \frac{\theta}{\sigma}}  p_{sig}'\bigg(z + \frac{\theta}{\sigma} \bigg) \phi(z) dz +
\int_{1.96 + \frac{\theta}{\sigma} }^{\infty}  p_{sig}'\bigg(z - \frac{\theta}{\sigma} \bigg) \phi(z) dz
\Bigg]
=
(\eta_1 + \eta_2) \theta
\end{equation}

In the second line, the second term in the sum equals zero because symmetry of $p_{sig}()$ and $\phi()$ about zero implies that both terms in the brackets are equal. To see why the third term in the sum equals zero, note that

\begin{equation}
\int_{-\infty}^{-1.96 - \frac{\theta}{\sigma}}  p_{sig}'\bigg(z + \frac{\theta}{\sigma} \bigg) \phi(z) dz
=
\int_{1.96 + \frac{\theta}{\sigma}}^{\infty}  p_{sig}'\bigg(-u + \frac{\theta}{\sigma} \bigg) \phi(u) du
=
-\int_{1.96 + \frac{\theta}{\sigma}}^{\infty}  p_{sig}'\bigg(u - \frac{\theta}{\sigma} \bigg) \phi(u) du
\end{equation}

The first equality follows from both changing the order of the integral limits and applying the substitution $u = -x$; it also uses the symmetry of $\phi()$. The final equality holds because symmetry of $p_{sig}()$ about zero implies that for any $t>1.96$, $p_{sig}'(t) = -p_{sig}'(-t)$.

\subsubsection*{Case 2.} Consider the second case where $\theta \geq 1.96\sigma$. For a given $(\theta,\sigma)$, we have

$$
\mathbbm{E}(X | 1.96\sigma \leq |X| )
=
\frac{1}{\bar{\eta}} \Bigg(
\eta_1 \mathbbm{E}(X | X \leq -1.96 \sigma)
+ \eta_2 \mathbbm{E}(X | X \geq 2\theta + 1.96\sigma )
$$
$$
\eta_3 \mathbbm{E}(X | 1.96 \sigma \leq X \leq  2\theta - 1.96\sigma )
+ \big(\bar{\eta}- \eta_1 - \eta_2 - \eta_3 \big) \mathbbm{E}(X | 2\theta - 1.96\sigma \leq X  \leq 2\theta + 1.96\sigma )
\Bigg)
$$
\begin{equation}
>
\frac{1}{\bar{\eta}} \Bigg(
\theta(\eta_1 + \eta_2)
+ \big(\bar{\eta}- \eta_1 - \eta_2 - \eta_3 \big) \theta
+ \eta_3 \mathbbm{E}(X | 1.96 \sigma \leq X \leq  2\theta - 1.96\sigma )
\Bigg)
\end{equation}

The inequality follows from two facts. First, the inequality proved in the first case: $\eta_1 \mathbbm{E}(X | X \leq -1.96 \sigma)
+ \eta_2 \mathbbm{E}(X | X \geq 2\theta + 1.96\sigma )\geq (\eta_1 + \eta_2) \theta$. Second, the expectation in the third term of the sum satisfies $\mathbbm{E}(X | 2\theta - 1.96\sigma \leq X  \leq 2\theta + 1.96\sigma ) >\theta$ because $\theta \geq 1.96\sigma \iff 2\theta - 1.96\sigma \geq \theta$ and we assume that $p_{sig}()>0$. 

It remains to show that $\mathbbm{E}(X | 1.96 \sigma \leq X \leq  2\theta - 1.96\sigma ) \geq \theta$. Then it follows that $\mathbbm{E}(X | 1.96\sigma \leq |X| )> \theta$, which is what we want to show. First, note that

\begin{equation}
\mathbbm{E}(X | 1.96 \sigma \leq X \leq  2\theta - 1.96\sigma )
= \theta + \sigma \mathbbm{E}\bigg(Z \bigg| 1.96 -\frac{\theta}{\sigma} \leq Z \leq  - 1.96 + \frac{\theta}{\sigma} \bigg)
\end{equation}

It is therefore sufficient to show that $\mathbbm{E}\bigg(Z \bigg| 1.96 -\frac{\theta}{\sigma} \leq Z \leq  - 1.96 + \frac{\theta}{\sigma} \bigg) \geq 0$. Writing out the expectation in full gives

$$
\mathbbm{E}\bigg(Z \bigg| 1.96 -\frac{\theta}{\sigma} \leq Z \leq  - 1.96 + \frac{\theta}{\sigma} \bigg)
=
\frac{1}{\eta_3} \Bigg(
\int_{1.96-\frac{\theta}{\sigma}}^{ 0 } z p_{sig} \bigg(z + \frac{\theta}{\sigma} \bigg) \phi(z) dz
+
\int_{0}^{\frac{\theta}{\sigma} - 1.96} z p_{sig} \bigg(z + \frac{\theta}{\sigma} \bigg) \phi(z) dz
\Bigg)
$$
\begin{equation}
=
\frac{1}{\eta_3} \Bigg(
\int_{0}^{\frac{\theta}{\sigma} - 1.96} 
z \bigg[
p_{sig} \bigg(z + \frac{\theta}{\sigma} \bigg)
 - p_{sig} \bigg(-z + \frac{\theta}{\sigma} \bigg)
\bigg]
\phi(z) dz
\Bigg)
\geq
0
\label{eq:prop2_final}
\end{equation}
\noindent

The second equality follows because
\begin{equation}
\int_{1.96-\frac{\theta}{\sigma}}^{ 0 } z p_{sig} \bigg(z + \frac{\theta}{\sigma} \bigg) \phi(z) dz
=
- \int_{0}^{1.96-\frac{\theta}{\sigma}} z p_{sig} \bigg(z + \frac{\theta}{\sigma} \bigg) \phi(z) dz
=
- \int_{0}^{\frac{\theta}{\sigma} - 1.96} u p_{sig} \bigg(-u + \frac{\theta}{\sigma} \bigg) \phi(u)du
\end{equation}

\noindent
which uses the substitution $u=-x$ and the symmetry of $\phi()$. The weak inequality in equation \eqref{eq:prop2_final} follows because $p_{sig}()$ is assumed to be weakly increasing over positive values. Thus, $z - \theta/\sigma > -z + \theta/\sigma$ for all $z \in (0, \theta/\sigma - 1.96)$ implies $p_{sig} \big(z + \theta/\sigma \big) - p_{sig} \big(-z + \theta \sigma \big) \geq 0$.

This covers all cases and proves the proposition.
 $\square$ \\

\newpage
\noindent
\textbf{Proof of Proposition 3:} For notational convenience, let $(X_{sig}, \Sigma_{sig}, \Theta_{sig})$ denote the distribution of latent studies $(X^*, \Sigma^*, \Theta^*)$ conditional on being published $(D=1)$ and statistically significant $(|X^*/\Sigma^*| \geq 1.96)$. The expected replication probability (Definition 2) under the common power rule (Definition 3) can be written as

$$
    \mathbbm{E}_{X^*, \Sigma^*, \Theta^* |D, R, S_X^*} \Big[ RP\Big( X^*, \Theta^*, \sigma_r(X^*, \beta^n) \Big) \Big| D=1, R=1, |X^*/\Sigma^*| \geq 1.96 \Big]
$$
$$
=
    \mathbbm{E}_{X, \Sigma, \Theta | S_X} \Big[
    RP\big( 
    X, \Theta, \sigma_r(X, \Sigma, \beta^n) 
    \big) 
    \big| |X/\Sigma| \geq 1.96 \Big]
$$
$$
    =
    \mathbbm{E}_{X_{sig}, \Sigma_{sig}, \Theta_{sig}} \Big[ RP\Big( X_{sig}, \Theta_{sig}, \sigma_r(X_{sig}, \beta^n) \Big) \Big]
$$
\begin{equation}
    =
    \mathbbm{E}_{\Sigma_{sig}, \Theta_{sig}} \Bigg[
    \mathbbm{E}_{X_{sig} | \Sigma_{sig}, \Theta_{sig}} \Big[ RP\Big( X_{sig}, \Theta_{sig}, \sigma_r(X_{sig}, \beta^n)\Big) | \Theta_{sig} = \theta, \Sigma_{sig} = \sigma \Big] 
    \Bigg]
    \label{eq:unconditional_rr}
\end{equation}

\noindent
where the second inequality drops the conditioning on being chosen for replication ($R$) because it is assumed that replication selection on significant results is random; and the last equality uses the Law of Iterated Expectations. The proof shows that the conditional expected replication probability satisfies $\mathbbm{E}_{X_{sig} | \Sigma_{sig}, \Theta_{sig}} \big[ RP\big( X_{sig}, \Theta_{sig}, \sigma_r(X_{sig}, \beta^n)\big) | \Theta_{sig} = \theta, \Sigma_{sig} = \sigma \big]  < 1 - \beta^n$ which implies that the expected replication probability is also less than intended power $1-\beta^n$. For greater clarity in what follows, let $\mathbbm{E} \big[ RP\big( X_{sig} | \theta, \sigma, \beta^n) \big]$ be shorthand for $\mathbbm{E}_{X_{sig} | \Sigma_{sig}, \Theta_{sig}} \big[ RP\big( X_{sig}, \Theta_{sig}, \sigma_r(X_{sig}, \beta^n)\big) | \Theta_{sig} = \theta, \Sigma_{sig} = \sigma \big]$.

Note that the conditional expected replication probability can be written explicitly as
\footnotesize
\begin{equation}
    \mathbbm{E} \big[ RP\big( X_{sig} | \theta, \sigma, \beta^n) \big]
    =
    \int \bigg( 1 -  \Phi \bigg( 1.96 - \frac{\theta}{x} \big(1.96 - \Phi^{-1}(\beta^n) \big) \bigg)
    \frac{ p\big(\frac{x}{\sigma} \big)\frac{1}{\sigma} \phi 
    \Big(\frac{x - \theta}{\sigma} \Big) \mathbbm{1}\big(|\frac{x}{\sigma} |\geq 1.96\big)dx
    }
    { 
    \int_{x'} p\big(\frac{x'}{\sigma})
    \frac{1}{\sigma} \phi \Big(\frac{x' - \theta}{\sigma} \Big) \mathbbm{1}\big(|\frac{x}{\sigma} |\geq 1.96\big) dx' 
    }
    \label{eq:conditional_rr}
\end{equation}
\normalsize
where the integrand in equation \eqref{eq:conditional_rr} is obtained using the compact notation for the replication probability derived in Lemma 2.1 and then substituting the common power rule in Definition 3. The pdf of estimates differs from a normal distribution in two respects: (1) the publication probability function $p\big(\frac{x}{\sigma} \big)$ reweights the distribution; and (2) conditioning on statistical significance truncates original effects falling in the insignificant region $(-1.96\sigma, 1.96\sigma)$. The denominator is the normalization constant.

First, we introduce some notation. Lemma 2.3 shows that if $(1-\beta^n)>0.6628$, then $RP\big( x, | \theta, \sigma, \beta^n \big)$ is strictly concave over the open interval $( \max{\{0, [1-r^*(\beta^n)]\theta}\}, [1+r^*(\beta^n)]\theta)$, where $r^*(\beta^n)$ is given by equation \eqref{eq:app_a_quad_solution}. This Proposition assumes $(1-\beta^n)>0.8314$, so the condition is satisfied. To simplify the notation, define $(l^*, u^*) = \big( (1-r^*)\theta, (1+r^*) \big)$ when $r^* \in (0,1)$ and $(l^*, u^*) = \big( 0, 2\theta \big)$ when $r^*\geq 1$; in both cases, the replication probability function is strictly concave over an interval with mid-point $\theta$.

Consider first the case where $r^*\geq 1$ so that $(l^*, u^*) = \big( 0, 2\theta \big)$. The conditional replication probability can be expressed as a weighted sum

\small
$$
    \mathbbm{E} \Big[ \big( RP\big( X_{sig} |\theta, \sigma, \beta^n \big) \Big]
    =
    \mathbbm{P}\Big( X_{sig} < l^* \Big) \mathbbm{E} \Big[ RP\big( X_{sig} |\theta, \sigma, \beta^n \big) \Big| X_{sig} <l^*\Big]
$$
$$
    +
    \mathbbm{P}\Big(l^* \leq X_{sig} \leq u^* \Big) \mathbbm{E} \Big[ RP\big( X_{sig} |\theta, \sigma, \beta^n\big) \Big|  l^* \leq X_{sig}  \leq u^*  \Big]
    +
    \mathbbm{P}\Big( X_{sig}  > u^* \Big) \mathbbm{E} \Big[ RP\big( X_{sig} |\theta, \sigma, \beta^n\big) \Big|  X_{sig}  > u^* \Big]
$$
\begin{equation}
    <
    \mathbbm{P}\Big( X_{sig}  < l^* \Big) 0.025
    +
    \mathbbm{P}\Big(l^* \leq X_{sig}  \leq u^* \Big) \mathbbm{E}  \Big[ RP\big( X_{sig} |\theta, \sigma, \beta^n\big) \Big|  l^* \leq X_{sig}  \leq u^*  \Big]
    +
    \mathbbm{P}\Big( X_{sig} > u^* \Big) \big( 1 - \beta^n \big)
    \label{eq:prop3_inequality_step1}
\end{equation}
\normalsize

In the last line, the first term in the sum uses the fact that the maximum value of the replication probability when $x<l^* = 0$ is 0.025 (Lemma 2.2 and Lemma 2.4 in Appendix A). The third term follows because $RP\big( 2\theta |\theta, \sigma, \beta^n\big)$ is the maximum value the function takes over $x > u^* = 2\theta$, since the function is strictly decreasing over $x>0$ (Lemma 2.2); and therefore that $RP\big( 2\theta |\theta, \sigma, \beta^n\big) < RP\big( \theta |\theta, \sigma, \beta^n\big) = 1-\beta^n$, where the equality is shown in Lemma 1. From equation \eqref{eq:prop3_inequality_step1}, we can see that $\mathbbm{E} \big[ RP ( X_{sig}|\theta, \sigma, \beta^n ) |  l^* \leq X_{sig} \leq u^* \big] < 1 - \beta^n$ is a sufficient condition for $\mathbbm{E} \big[ RP\big( X_{sig}|\theta, \sigma, \beta^n) \big] < 1-\beta^n$. 

Before showing that this sufficient condition is satisfied, we show that the same sufficient condition holds in the second case, where $r^* \in (0,1)$ so that $(l^*, u^*) = \big( (1-r^*)\theta, (1+r^*)\theta \big)$. This requires additional steps. First, express the conditional replication probability as a weighted sum
\small
$$
    \mathbbm{E} \Big[ \big( RP\big( X_{sig}|\theta, \sigma, \beta^n \big) \Big]
    =
    \mathbbm{P}\Big( X_{sig} \leq l^* \Big) \mathbbm{E} \Big[ RP\big( X_{sig}|\theta, \sigma, \beta^n \big) \Big| X_{sig} \leq l^* \Big]
$$
$$
    +
    \mathbbm{P}\Big(l^* \leq X_{sig} \leq u^* \Big) \mathbbm{E} \Big[ RP\big( X_{sig}|\theta, \sigma, \beta^n\big) \Big|  l^* \leq X_{sig} \leq u^*  \Big]
    +
    \mathbbm{P}\Big( X_{sig} \geq u^* \Big) \mathbbm{E} \Big[ RP\big( X_{sig}|\theta, \sigma, \beta^n\big) \Big|  X_{sig} \geq u^* \Big]
$$
\begin{equation}
< 
\mathbbm{P}\Big( X_{sig} \leq l^* \Big)
+
\mathbbm{P}\Big( l^* \leq X_{sig} \leq u^*  \Big) \mathbbm{E} \Big[ RP\big( X_{sig}|\theta, \sigma, \beta^n\big) \Big|  l^* \leq X_{sig} \leq u^* \Big]
+ 
\mathbbm{P}\Big( X_{sig} \geq u^* \Big) RP\Big(u^* | \theta, \sigma, \beta^n \Big)
\end{equation}
\normalsize

The strict inequality follows for two reasons. For the first term in the sum, one is the maximum value the function can take for any $x$. For the third term, $RP(u^* | \theta, \sigma, \beta^n)$ is the function's maximum value over $x\geq u^*$, since the integrand is strictly decreasing over positive values (Lemma 2.2). With an additional step, we can write this inequality as

$$
\mathbbm{E} \Big[ \big( RP\big( X_{sig}|\theta, \sigma, \beta^n \big) \Big]
<
\frac{1}{2} \Big( 1 - \mathbbm{P}\Big( l^* \leq X_{sig} \leq u^* \Big) \Big)
\Big( 1 + RP\big( u^* \big| \theta, \sigma, \beta^n \big) \Big)
$$
\begin{equation}
+
\mathbbm{P}\Big( l^*\leq X_{sig} \leq u^*  \Big) \mathbbm{E} \Big[ RP\big( X_{sig}|\theta, \sigma, \beta^n\big) \Big| l^* \leq X_{sig} \leq u^* \Big]
\label{eq:prop3_inequality_step2}
\end{equation}

This follows because $\mathbbm{P}( X_{sig} \leq l^*) \leq \mathbbm{P} ( X_{sig} \geq u^* )$ and $RP( u^* | \theta, \sigma, \beta^n ) < 1$. That is, increasing the relative weight on the maximum value of one, such that both tails are equally weighted, must lead to a (weakly) larger value. The weak inequality $\mathbbm{P}( X_{sig} \leq l^*) \leq \mathbbm{P} ( X_{sig} \geq u^* )$ required for this simplification is shown below: \\

\noindent 
\textbf{Lemma 3.} \textit{Suppose $X| \theta, \sigma$ follows the truncated normal pdf in equation \eqref{eq:conditional_rr}. Then for any $r^* \in (0,1)$, the following inequality holds: $\mathbbm{P}\big( X_{sig} \leq (1-r^*)\theta \big) < \mathbbm{P}\big( X_{sig} \geq (1+r^*)\theta \big)$.}

\begin{proof}
First, note that $\big( (1-r^*)\theta, (1+r^*)\theta \big)$ is an interval over the positive real line centered at $\theta$. Consider two cases: \\

\textit{Case 1:} Let $(1-r^*)\theta \leq 1.96 \sigma$. Define the normalization constant $C = \int_{x'} p\big(\frac{x'}{\sigma})
    \frac{1}{\sigma} \phi \Big(\frac{x' - \theta}{\sigma} \Big) \mathbbm{1}\big( | \frac{x}{\sigma} |\geq 1.96\big) dx'$. Then

\small
$$
\mathbbm{P}\Big( X_{sig} \leq (1-r^*)\theta \Big) 
=
\frac{1}{C} \int_{-\infty}^{-1.96\sigma} p_{sig}\bigg(\frac{x}{\sigma} \bigg)
    \frac{1}{\sigma} \phi \bigg(\frac{x - \theta}{\sigma} \bigg)  dx' 
\leq 
\frac{1}{C} \int_{2\theta + 1.96 \sigma}^{\infty} p_{sig}\bigg(\frac{x}{\sigma} \bigg)
    \frac{1}{\sigma} \phi \bigg(\frac{x - \theta}{\sigma} \bigg)  dx' 
$$
\begin{equation}
<
\frac{1}{C} \int_{2\theta + 1.96 \sigma}^{\infty} p_{sig}\bigg(\frac{x}{\sigma} \bigg)
    \frac{1}{\sigma} \phi \bigg(\frac{x - \theta}{\sigma} \bigg)  dx' 
+
\frac{1}{C} \int_{\max{\{1.96\sigma, (1+r^*)\theta\}}}^{2\theta + 1.96 \sigma} p_{sig}\bigg(\frac{x}{\sigma} \bigg)
    \frac{1}{\sigma} \phi \bigg(\frac{x - \theta}{\sigma} \bigg)  dx' 
    = \mathbbm{P}\Big( X_{sig} \geq (1+r^*)\theta \Big) 
\end{equation}
\normalsize

Consider the weak inequality. Note that the mid-point between $-1.96\sigma$ and $2\theta + 1.96\sigma$ is $\theta$. Thus, with no publication bias (i.e. $p(t) = 1$ for all $t$), we would have equality owing to the symmetry of the normal distribution. However, recall that $p_{sig}()$ is symmetric about zero and weakly increasing in absolute value. It follows therefore that $|2\theta + 1.96\sigma |> |-1.96\sigma|$ implies $p_{sig}(|2\theta + 1.96\sigma |) \geq p_{sig}(|-1.96\sigma|)$; using this fact and symmetry of the normal distribution about $\theta$ gives the weak inequality. The strict inequality follows because the additional term is strictly positive, since $p_{sig}()$ is assumed to be non-zero. \\

\textit{Case 2:} Let $(1-r^*)\theta > 1.96 \sigma$. The argument is similar to the first case:
\small
$$
\mathbbm{P}\Big( X_{sig} \leq (1-r^*)\theta \Big) 
=
\frac{1}{C} \int_{-\infty}^{-1.96\sigma} p_{sig}\bigg(\frac{x}{\sigma} \bigg)
    \frac{1}{\sigma} \phi \bigg(\frac{x - \theta}{\sigma} \bigg)  dx' 
+
\frac{1}{C} \int_{1.96\sigma}^{(1-r^*)\theta} p_{sig}\bigg(\frac{x}{\sigma} \bigg)
    \frac{1}{\sigma} \phi \bigg(\frac{x - \theta}{\sigma} \bigg)  dx' 
$$
$$
<
\frac{1}{C} \int_{2\theta + 1.96 \sigma}^{\infty} p_{sig}\bigg(\frac{x}{\sigma} \bigg)
    \frac{1}{\sigma} \phi \bigg(\frac{x - \theta}{\sigma} \bigg)  dx' 
+
\frac{1}{C} \int_{(1+r^*)\theta}^{2\theta - 1.96 \sigma} p_{sig}\bigg(\frac{x}{\sigma} \bigg)
    \frac{1}{\sigma} \phi \bigg(\frac{x - \theta}{\sigma} \bigg)  dx'
$$
\begin{equation}
+
\frac{1}{C} \int_{2\theta - 1.96 \sigma}^{2\theta + 1.96 \sigma} p_{sig}\bigg(\frac{x}{\sigma} \bigg)
    \frac{1}{\sigma} \phi \bigg(\frac{x - \theta}{\sigma} \bigg)  dx'
    = \mathbbm{P}\Big( X_{sig} \geq (1+r^*)\theta \Big) 
\end{equation}
\normalsize
\end{proof}

The inequality in equation \eqref{eq:prop3_inequality_step2} can be further simplified by placing restrictions on intended power. In particular, if intended power satisfies $1-\beta^n \geq 0.8314$, then

$$
\mathbbm{E} \Big[ \big( RP\big( X_{sig}|\theta, \sigma, \beta^n \big) \Big]
<
\Big( 1 - \mathbbm{P}\big( l^*  \leq X_{sig} \leq u^*  \big) \Big)
\big( 1 - \beta^n \big)
$$
\begin{equation}
+
\mathbbm{P}\big( l^* \leq X_{sig} \leq u^*  \big) \mathbbm{E} \Big[ RP\big( X_{sig}|\theta, \sigma, \beta^n\big) \Big|  l^* \leq X_{sig} \leq u^* \Big]
\label{eq:prop3_inequality_step3}
\end{equation}

This follows because with $u^* = (1+r^*)\theta$, we have

$$
\frac{1}{2} \Big( 1 + RP\big( u^* \big| \theta, \sigma, \beta^n \big)  \Big)
=
\frac{1}{2} 
\Bigg(1 + 
\Bigg( 1 -\Phi \Bigg(1.96 - \frac{1.96 - \Phi^{-1}(\beta^n)}{1 + r^*(\beta^n)} \Bigg)
\Bigg)
$$
\begin{equation}
\leq 1 - \beta^n \iff 1-\beta^n \geq 0.8314
\end{equation}

From equation \eqref{eq:prop3_inequality_step3}, we can see that $\mathbbm{E} \big[ RP ( X_{sig}|\theta, \sigma, \beta^n ) \big|  l^* \leq X_{sig} \leq u^* \big] < 1 - \beta^n$ is a sufficient condition for $\mathbbm{E} \big[ RP( X_{sig}|\theta, \sigma, \beta^n) \big] < 1-\beta^n$. Thus, in both cases, the sufficient condition for the desired result is the same.

This sufficient condition is shown in two steps. In the first, I show that this inequality holds even in the case where there is no publication bias and all published results are replicated (i.e. when $X \sim N(\Theta, \Sigma^2)$). In the second, I show that this inequality remains true once we allow for publication bias and truncation of the distribution due to conditioning on statistical significance.

Lemma 4 states the first intermediate step. Its implications are of independent interest and discussed in the main text. It shows that even in the optimistic scenario where original estimates are unbiased, there is no publication bias, and all results are published and replicated, that the expected replication probability still falls below intended power. \\

\textbf{Lemma 4.} \textit{Let published effects be distributed according to $X| \theta, \sigma \sim N(\theta, \sigma^2)$. Suppose $p(t)=1$ and $r(t)=1$ for all $t \in \mathbb{R}$. Assume all results are included in the replication rate calculation. Let power in replications is set according to the common power rule with intended power $1-\beta^n \geq 0.8314$.  Then $\mathbbm{E} \big[ RP( X |\theta, \sigma, \beta^n \big) \big] < 1-\beta^n$.}

\begin{proof}
Recall that $RP(x | \theta, \sigma, \beta^n)$ is strictly concave with respect to $x$ over the interval $(l^*, u^*)$, where $(l^*, u^*) = \big( (1-r^*)\theta, (1+r^*) \big)$ when $r^* \in (0,1)$ and $(l^*, u^*) = \big( 0, 2\theta \big)$; in both cases, the mid-point of the interval is $\theta$. We have that

\small
\begin{equation}
    \mathbbm{E} \Big[ RP\big( X|\theta, \sigma, \beta^n\big) \Big|  l^* \leq X \leq u^* \Big]
    =
    \bigintssss_{l^*}^{u^*}
    RP\big( x |\theta, \sigma, \beta^n\big)
    \frac{ \frac{1}{\sigma} \phi 
    \Big(\frac{x - \theta}{\sigma} \Big)dx
    }
    { 
    \int_{l^*}^{u^*}
    \frac{1}{\sigma} \phi \Big(\frac{x' - \theta}{\sigma} \Big)dx' 
    }
    <
    RP\Big( \theta \Big| \theta, \sigma, \beta^n\big)  \Big) 
    =
    1-\beta^n
\end{equation}
\normalsize
\noindent
where the strict inequality follows from Jensen's inequality and the fact that $\mathbbm{E}[X | l^* \leq X \leq u^*]=\theta$. The final equality is a property of the replication probability function shown in Lemma 1. This is the sufficient condition required for the desired result.

Note that the inequalities in equations \eqref{eq:prop3_inequality_step2} (for when $r^* \geq1 )$ and \eqref{eq:prop3_inequality_step3} (for when $r^* \in (0,1)$) were derived under more general conditions, where the normal distribution may we reweighted by $p()$ and truncated based on significance. This setting is a special case with no publication bias (i.e. $p(t)=1$ for all $t$), and no truncation such that all results are included in the replication rate irrespective of statistical significance.
\end{proof}

The same conclusions hold when we introduce publication bias (which reweights the normal distribution) and condition on statistical significance (which truncates the `insignificant' regions of the density). Consider three cases. First, suppose that $u^* \leq 1.96\sigma$. Then $\mathbbm{E} \big( RP\big( X_{sig}|\theta, \sigma, \beta^n\big) \Big|  l^* \leq X_{sig} \leq u^* \big) = 0 < 1-\beta^n$ because of truncation. Second, suppose that $l^* \geq 1.96\sigma$. Then 

$$
    \mathbbm{E} \Big[ RP\big( X_{sig}|\theta, \sigma, \beta^n\big) \Big|  l^* \leq X_{sig} \leq u^* \Big]
    =
    \bigintssss_{l^*}^{u^*}
    RP\big( x |\theta, \sigma, \beta^n\big)
    \frac{ p_{sig} \big(\frac{x}{\sigma} \big) \frac{1}{\sigma} \phi 
    \Big(\frac{x - \theta}{\sigma} \Big)  dx
    }
    { 
    \int_{l^*}^{u^*}
    p_{sig} \big(\frac{x}{\sigma} \big)
    \frac{1}{\sigma} \phi \Big(\frac{x' - \theta}{\sigma} \Big)  dx' 
    }
$$
\begin{equation}
    \leq
    \bigintssss_{l^*}^{u^*}
    RP\big( x |\theta, \sigma, \beta^n\big)
    \frac{ \frac{1}{\sigma} \phi 
    \Big(\frac{x - \theta}{\sigma} \Big)dx
    }
    { 
    \int_{l^*}^{u^*}
    \frac{1}{\sigma} \phi \Big(\frac{x' - \theta}{\sigma} \Big)dx' 
    }
    <
    RP\Big( \theta \Big| \theta, \sigma, \beta^n\big)  \Big) 
    =
    1-\beta^n
\end{equation}

Note that the distribution is invariant to the scale of $p_{sig}()$. Consider first the weak inequality. This follows because $p_{sig}()$ is assumed to be weakly increasing over $(l^*, u^*)$. When it is a constant function over the interval, the equality holds. If $p_{sig}(x/\sigma)>0$ for some $x \in (l^*, u^*)$ then the function redistributes weight to larger values of $x$. Since $RP( x |\theta, \sigma, \beta^n )$ is strictly decreasing over positive values of $x$ (Lemma 2.2 in Appendix A), placing higher relative weight on lower values implies that the weak inequality becomes strict. As in the proof to Lemma 4, the strict inequality follows from Jensen's inequality, since $RP( x |\theta, \sigma, \beta^n )$ is strictly concave over $(l^*, u^*)$, and the fact that the expected value of $X$ over this interval is equal to the true value $\theta$. The last equality follows from Lemma 1 in the text.

Finally, consider the case where $l^* < 1.96\sigma < u^*$. Then

\footnotesize
$$
    \mathbbm{E} \Big[ RP\big( X_{sig}|\theta, \sigma, \beta^n\big) \Big|  l^* \leq X_{sig} \leq u^* \Big]
    =
    \bigintssss_{1.96\sigma}^{u^*}
    RP\big( x |\theta, \sigma, \beta^n\big)
    \frac{ p_{sig} \big(\frac{x}{\sigma} \big) \frac{1}{\sigma} \phi 
    \Big(\frac{x - \theta}{\sigma} \Big)  dx
    }
    { 
    \int_{1.96\sigma}^{u^*}
    p_{sig} \big(\frac{x'}{\sigma} \big)
    \frac{1}{\sigma} \phi \Big(\frac{x' - \theta}{\sigma} \Big)  dx' 
    }
$$
$$
    =
    \bigintssss_{1.96\sigma}^{2\theta - 1.96\sigma}
    RP\big( x |\theta, \sigma, \beta^n\big)
    \frac{ p_{sig} \big(\frac{x}{\sigma} \big) \frac{1}{\sigma} \phi 
    \Big(\frac{x - \theta}{\sigma} \Big)  dx
    }
    { 
    \int_{1.96\sigma}^{u^*}
    p_{sig} \big(\frac{x'}{\sigma} \big)
    \frac{1}{\sigma} \phi \Big(\frac{x' - \theta}{\sigma} \Big)  dx' 
    }
    +
    \bigintssss_{2\theta - 1.96\sigma}^{u^*}
    RP\big( x |\theta, \sigma, \beta^n\big)
    \frac{ p_{sig} \big(\frac{x}{\sigma} \big) \frac{1}{\sigma} \phi 
    \Big(\frac{x - \theta}{\sigma} \Big)  dx
    }
    { 
    \int_{1.96\sigma}^{u^*}
    p_{sig} \big(\frac{x'}{\sigma} \big)
    \frac{1}{\sigma} \phi \Big(\frac{x' - \theta}{\sigma} \Big)  dx' 
    }
$$
$$
=
    \omega
    \bigintssss_{1.96\sigma}^{2\theta - 1.96\sigma}
    RP\big( x |\theta, \sigma, \beta^n\big)
    \frac{ p_{sig} \big(\frac{x}{\sigma} \big) \frac{1}{\sigma} \phi 
    \Big(\frac{x - \theta}{\sigma} \Big)  dx
    }
    { 
    \int_{1.96\sigma}^{2\theta - 1.96\sigma}
    p_{sig} \big(\frac{x'}{\sigma} \big)
    \frac{1}{\sigma} \phi \Big(\frac{x' - \theta}{\sigma} \Big)  dx' 
    }
    +
    (1-\omega)
    \bigintssss_{2\theta - 1.96\sigma}^{u^*}
    RP\big( x |\theta, \sigma, \beta^n\big)
    \frac{ p_{sig} \big(\frac{x}{\sigma} \big) \frac{1}{\sigma} \phi 
    \Big(\frac{x - \theta}{\sigma} \Big)  dx
    }
    { 
    \int_{2\theta - 1.96\sigma}^{u^*}
    p_{sig} \big(\frac{x'}{\sigma} \big)
    \frac{1}{\sigma} \phi \Big(\frac{x' - \theta}{\sigma} \Big)  dx' 
    }
$$
$$
=
    \omega
    \bigintssss_{1.96\sigma}^{2\theta - 1.96\sigma}
    RP\big( x |\theta, \sigma, \beta^n\big)
    \frac{ \frac{1}{\sigma} \phi 
    \Big(\frac{x - \theta}{\sigma} \Big)  dx
    }
    { 
    \int_{1.96\sigma}^{2\theta - 1.96\sigma}
    \frac{1}{\sigma} \phi \Big(\frac{x' - \theta}{\sigma} \Big)  dx' 
    }
    +
    (1-\omega)
    \bigintssss_{2\theta - 1.96\sigma}^{u^*}
    RP\big( x |\theta, \sigma, \beta^n\big)
    \frac{ \frac{1}{\sigma} \phi 
    \Big(\frac{x - \theta}{\sigma} \Big)  dx
    }
    { 
    \int_{2\theta - 1.96\sigma}^{u^*}
    \frac{1}{\sigma} \phi \Big(\frac{x' - \theta}{\sigma} \Big)  dx' 
    }
$$
\begin{equation}
    <
    \omega     RP\Big( \theta \Big| \theta, \sigma, \beta^n\big)  \Big) 
    +
    (1-\omega). RP\Big( 2\theta - 1.96\sigma \Big| \theta, \sigma, \beta^n\big)  \Big) < 1-\beta^n
\end{equation}
\normalsize

\noindent
with 
\begin{equation}
\omega = 
\frac{
    \int_{1.96\sigma}^{2\theta - 1.96\sigma}
    p_{sig} \big(\frac{x'}{\sigma} \big)
    \frac{1}{\sigma} \phi \Big(\frac{x' - \theta}{\sigma} \Big)  dx' 
}
{
    \int_{1.96\sigma}^{u^*}
    p_{sig} \big(\frac{x'}{\sigma} \big)
    \frac{1}{\sigma} \phi \Big(\frac{x' - \theta}{\sigma} \Big)  dx' 
}
\end{equation}

The second row simply breaks up the integral. The third row rearranges the sum so that the conditional expectation of the replication probability appears in both terms. The third line follows because, as in the previous case, the $p_{sig}$ function redistributes weight to large values of $x$ and hence lower values of $RP(x|\theta, \sigma, \beta^n)$. In the last line, the first term uses the concavity of $RP(x|\theta, \sigma, \beta^n)$ over $(1.96\sigma, 2\theta - 1.96\sigma) \subset (l^*, u*)$, Jensen's inequality, and the fact that the expected value of $X$ over this interval is equal to $\theta$. The second term follows because $2\theta - 1.96\sigma$ is the maximum value the function can take because $RP(x|\theta, \sigma, \beta^n)$ is strictly decreasing in $x$ over positive values. The final inequality follows because $RP\big( \theta \big| \theta, \sigma, \beta^n\big)  \big) = 1 - \beta^n$ (Lemma 1) and $RP\big( 2\theta - 1.96\sigma \big| \theta, \sigma, \beta^n\big)  \big) < 1-\beta^n$ because $2\theta - 1.96\sigma > \theta$ and the function is strictly decreasing over positive values. 

This covers all cases, proving the proposition.

\newpage
\section*{C. Publication Bias Above 1.96 and the Replication Rate}

Proposition 1 shows that the replication rate does not depend on the probability of publishing insignificant results relative to significant results. However, it may vary with changes in $p_{sig}()$ i.e when the absolute value of the $t$-ratio is above 1.96. This section presents a simple example showing how the replication rate varies with the relative probability of publishing `moderately significant' results to `highly significant' results.

\begin{figure} [H]
	\centering
\includegraphics[width=1\textwidth]{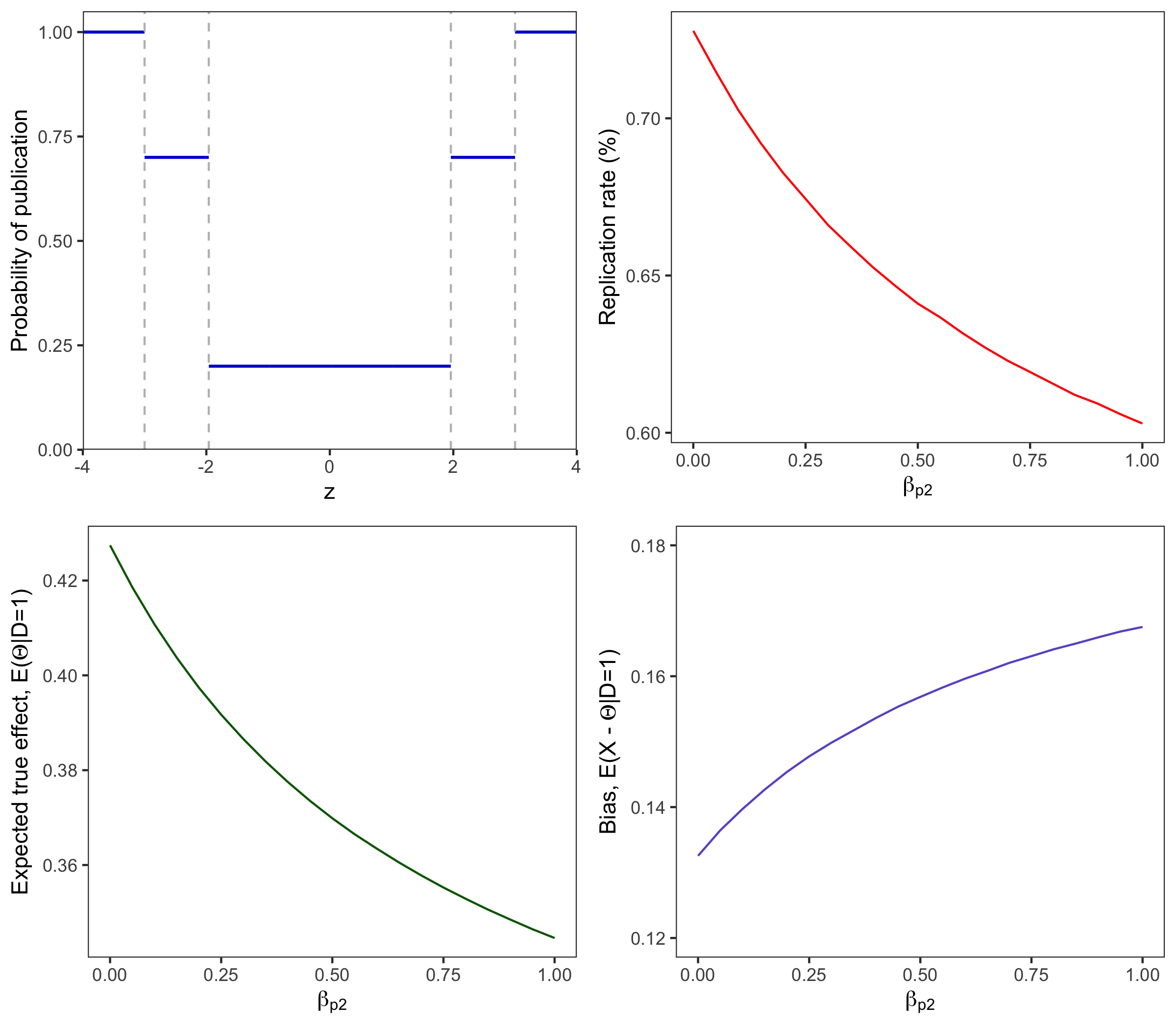} 
	\caption{The replication rate, true effect, and bias changing the probability of publishing moderately significant results relative to highly significant results ($\beta_{p2}$). Results are based on estimated parameters for economics experiments in Table \ref{tab:mle_results}, setting $\kappa = 3$ and varying $\beta_{p2}$. Power in replication studies is set detect the original estimate with 92\% power. The top left panel is an illustration of a stepwise publication probability function which distinguishes between insignificant findings $z \in (-1.96.1.96)$, moderately statistically significant findings $|z| \geq 1.96$ and $|z| \leq 3$, and highly statistically significant findings $|z|>3$. The top-right panel shows the replication rate, which is defined as the share of significant results that obtain a statistically significant results with the same sign in replication. The bottom-left panel plots the expected true value of published results and the bottom-right mean bias.}
	\label{fig:pb_above2_sim_92}
\end{figure}

Suppose that the probability function $p()$ is a stepwise function that distinguishes between insignificant findings, moderately statistically significant findings, and highly statistically significant findings. Specifically, let $\kappa > 1.96$ be a value such that $|z| \in (1.96, \kappa)$ is defined as moderately significant and $|z|>\kappa$ as highly significant. Let $\beta_{p1}$ refer to the constant probability of publishing insignificant findings, and $\beta_{p2}$ to the constant probability of publishing moderately significant findings. Both these probabilities are defined relative to the probability of publishing a highly significant finding, which we normalize to 1 (since only the ratio of probabilities are identified). The top-left panel of Figure \ref{fig:pb_above2_sim_92} provides an illustration with $\kappa =3$, $\beta_{p2} = 0.7$ and $\beta_{p1} = 0.2$. 

The top-right panel of Figure \ref{fig:pb_above2_sim_92} shows how the replication rate in economics experiments varies with $\beta_{p2}$. The results show that as $\beta_{p2}$ decreases -- that is, as highly significant results become more favoured for publication relative to moderately significant results -- the replication rate increases. The size of the changes in the replication rate as we vary $\beta_{p2}$ can be relatively large, increasing, for example, by more than 10 percentage points when we move from $\beta_{p2} = 1$ to $\beta_{p2}=0$.

The intuition is that increasing the relative probability of publishing highly significant results compared to moderately significant results (i.e. decreasing $\beta_{p2}$) has the effect of increasing the mean true effect in original published studies (bottom-left panel of Figure \ref{fig:pb_above2_sim_92}). All else equal, increasing the mean true effect will increase power, and therefore the replication rate. This is because the larger the true effect, the smaller is the bias of the original estimate (bottom-right panel of Figure \ref{fig:pb_above2_sim_92}). Power will therefore increase with larger true effects based on the rule. While this may be counter-intuitive from the perspective of publication bias -- if we think of favouring highly significant results over moderately significant results as a `worsening' of publication bias -- it is in fact very similar to what \citet{Benjamin2018} propose for setting a new standard of statistical significance for novel findings at $p<0.005$. \citet{McShane2019} provide arguments against this proposal.

\newpage
\section*{D. Extending the Replication Rate Definition}
This appendix analyzes a generalization of the replication rate definition that extends to insignificant results. It outlines a number of issues with this proposal.

Suppose we extend the definition of the replication rate such that insignificant original results are counted as `successfully replicated' if they are also insignificant in replications. Assume replication selection is a random sample of published results. Then we have the following definitions: \\

\noindent
\textbf{Definition D.1} (Generalized replication probability of a single study). \textit{The replication probability of a study $(X, \Sigma, \Theta)$ which is published $(D=1)$ and chosen for replication ($R=1$) is}
\scriptsize
\begin{equation}
    \widetilde{RP}\Big( X, \Theta, \sigma_r(X, \Sigma, \beta^n) \Big) =
    \begin{cases} 
          \mathbbm{P} \bigg( \frac{| {X}_r | }{\sigma_r(X, \Sigma, \beta^n)} \geq 1.96 , \text{sign}(X) = \text{sign}(X_r) \Big| X,  \Theta, \sigma_r(X, \Sigma, \beta^n)  \bigg) & \text{if } 1.96. \Sigma \leq | X |  \\
          \mathbbm{P} \bigg( \frac{| {X}_r | }{\sigma_r(X, \Sigma, \beta^n)} < 1.96 
          \Big| X,  \Theta, \sigma_r(X, \Sigma, \beta^n)
          \bigg) & \text{if } 1.96. \Sigma > | X |  \\
    \end{cases}
\end{equation}
\normalsize

\noindent
\textbf{Definition D.2} (Expected generalized replication probability). \textit{The expected generalized replication probability equals}

\scriptsize
$$
    \mathbbm{E} \Big[ \widetilde{RP}\Big( X, \Theta, \sigma_r(X, \Sigma, \beta^n) \Big) \Big]
    =
    \mathbbm{P} \Big(  1.96. \Sigma \leq \big|X\big| \Big) 
    \mathbbm{E} \Bigg[ 
    \widetilde{RP}\Big( X, \Theta, \sigma_r(X, \Sigma, \beta^n)
    \Bigg| 
    X, \Theta, \sigma_r(X, \Sigma, \beta^n), 1.96. \Sigma \leq \big|X\big|
    \Bigg]
$$
\begin{equation}
    + \bigg(1 - \mathbbm{P} \Big(  1.96. \Sigma \leq \big|X\big| \Big) \bigg)
    \mathbbm{E} \Bigg[ 
    \widetilde{RP}\Big( X, \Theta, \sigma_r(X, \Sigma, \beta^n)
    \Bigg|
    X, \Theta, \sigma_r(X, \Sigma, \beta^n),
     1.96. \Sigma > \big|X\big|
    \Bigg]
\end{equation}
\normalsize

First, note that Definition D.2 equals the standard replication rate definition when the expectation is taken only over significant studies because, in this case, $\mathbbm{P} \big( |X | \geq 1.96. \Sigma  \big)=1$. The degree to which the expected generalized replication probability differs from the standard expected replication probability depends on two factors. First, the share of published results that are insignificant. Second, the expected probability that replications will be insignificant conditional on original estimates being insignificant; this in turn depends on the power rule implemented in replication studies.\footnote{Additionally, note that this definition implies that if $\theta = 0$, then $\widetilde{RP}\Big( X, \Theta, \sigma_r(X, \Sigma, \beta^n) | \Theta = 0 \Big) = 0.90375$. That is, the replication probability of null results is constant and independent of power in original studies and replication studies.}

To analyze the generalized replication rate, we can apply the empirical approach outlined in the main text, but using the generalized definition in place of the original definition. Recall that the original replication rate is invariant to publication bias against null results (Proposition 1). The generalized replication rate, by contrast, does vary as the degree of publication bias against null results changes. Two sets of results are therefore presented for comparison. The first set assumes publication bias using estimated selection parameters in Table \ref{tab:mle_results}. The second set assumes no publication bias (i.e. that all results are published with equal probability). We examine two rules for calculating replication power: the common power rule and the original power rule (where the replication standard error is set equal to the original standard error). For more details on different rules for calculation replication power, see Appendix F.

Table \ref{tab:extended_rr_sim_results} reports the results for both applications. Under the common power rule, the simulated generalized replication rate remains below intended power in both publication regimes. Under the original power rule, it is relatively low when there is publication bias and around 80\% when there is no publication bias. 

These generalized replication rate predictions differs from the standard replication rate predictions for two reasons: (i) the share of insignificant results in the published literature and (ii) the replication probability when results are insignificant, which depends on the power rule used in replication studies. 

On the first point, moving from the publication bias regime to the no publication bias regime implies a dramatic increase in the share of insignificant published results; in both applications, null results change from a minority of published results to a majority. 

On the second point, the results show that the replication power rules considered here have some undesirable properties. First, note that the common power rule is designed to detect original estimates with high statistical power. This implies that low-powered, insignificant original results will be high-powered in replications, which increases the probability that they are significant and thus counted as replication `failures' under the generalized definition. The original power rule has the reverse problem. On the one hand, low-powered, insignificant original studies are likely to be insignificant in replications, which counts as a `successful' replication under the generalized definition. However, on the other hand, low-powered, significant original studies will have low replication probabilities when the same low-powered design is repeated in replications. The generalized replication rate therefore depends crucially on the share of significant and insignificant findings in the published literature, and the distribution of standard errors. Under the original power rule with no publication bias, the generalized replication rate is around 80\% in both applications; however, with greater power in original studies, the replication rate would fall.

While the generalized replication rate changes as publication bias is reduced, the direction of this change depends on which replication power rule is used: with the original power rule the replication rate increases, while with the common power rule it decreases.

Overall, generalizing the replication rate with Definition D.2 does not deliver replication rates close to intended power under the common power rule. For the original power rule, it is higher when there is no publication bias because replications repeat low-power designs for low-powered original studies with insignificant results. The generalized replication rate under this original power rule will therefore be sensitive to the distribution of power in original studies.

\begin{table}[H] \centering 
  \normalsize
  \caption{\textsc{Simulated Generalized Replication Rate Results}}
\begin{tabularx}{0.6838\textwidth}{@{\extracolsep{5pt}} lcc} 
\\[-1.8ex]\hline 
\hline \\[-1.8ex] 
 &  \multicolumn{2}{c}{\textit{Simulated statistics}} \\ 
 & \\\cline{2-3}
\textbf{A Economics experiments}  & $\text{ }$92\% for $X$ & Original power  \\ 
\hline \\[-1.8ex] 
\underline{\textit{Publication bias                  }}      &   &   \\ 
Generalized replication rate               & 0.600 & 0.555   \\ 
$\mathbbm{P}(\text{Replicated} |S_X = 1)$  & 0.601 & 0.552   \\ 
$\mathbbm{P}(\text{Replicated} |S_X = 0)$  & 0.542 & 0.774  \\ 
$\mathbbm{P}(S_X = 1)$                     & 0.988 & 0.988   \\ 
$\mathbbm{P}(S_X = 0)$                     & 0.016 & 0.016   \\ 
&  & \\

\underline{\textit{No publication bias}}   &  &   \\ 
Generalized replication rate                 & 0.436  & 0.789 \\ 
$\mathbbm{P}(\text{Replicated} |S_X = 1)$   & 0.601 & 0.551   \\ 
$\mathbbm{P}(\text{Replicated} |S_X = 0)$   & 0.385 & 0.862  \\ 
$\mathbbm{P}(S_X = 1)$                      & 0.236 & 0.236   \\ 
$\mathbbm{P}(S_X = 0)$                      & 0.764 & 0.764   \\ 
&  & \\

\textbf{B Psychology experiments}    & &     \\ 
\hline \\[-1.8ex] 
\underline{\textit{Publication bias}}        &  &  \\ 
Generalized replication rate                 & 0.541 & 0.526  \\ 
$\mathbbm{P}(\text{Replicated} |S_X = 1)$   & 0.539 & 0.478   \\ 
$\mathbbm{P}(\text{Replicated} |S_X = 0)$   & 0.554 & 0.824    \\ 
$\mathbbm{P}(S_X = 1)$                      & 0.861  & 0.861    \\ 
$\mathbbm{P}(S_X = 0)$                      & 0.139 & 0.139   \\ 
& &  \\

\underline{\textit{No publication bias}}     &  &   \\ 
Generalized replication rate                & 0.474 & 0.805  \\ 
$\mathbbm{P}(\text{Replicated} |S_X = 1)$   & 0.537 & 0.479   \\ 
$\mathbbm{P}(\text{Replicated} |S_X = 0)$   & 0.460 & 0.88    \\ 
$\mathbbm{P}(S_X = 1)$                      & 0.188 & 0.188   \\ 
$\mathbbm{P}(S_X = 0)$                      & 0.812  & 0.812    \\ 
\hline
\hline
\end{tabularx} 
\caption*{\textit{Notes}: Economics experiments refer to \citet{Camerer2016} and psychology experiments to \citet{OpenScience2015}. The generalized replication rate is defined in the text. The indicator variable $S_X$ equals one for significant results and zero otherwise. Economics experiments refers to \citet{Camerer2016} and psychology experiments to \citet{OpenScience2015}. Simulated statistics are based on parameter estimates in Table \ref{tab:mle_results}. Different column represent different rules for calculating power in replications.}
\label{tab:extended_rr_sim_results}
\end{table}

\section*{E. Replication Selection in Empirical Applications}
Replication selection is a multi-step mechanism that first selects studies, and then selects results within those studies to replicate (since studies typically report multiple results). It consists of three steps:

\begin{enumerate}
    \item \textbf{Eligibility}: define the set of eligible studies (e.g. journals, time-frame, study designs).
    \item \textbf{Study selection:} on the set of eligible studies, a mechanism that select which studies will be included in the replication study.
    \item \textbf{Within-study replication selection:} for selected studies, a mechanism for selecting which result(s) to replicate.
\end{enumerate}

These three features of the replication selection mechanism determine: (i) the latent distribution estimated in the empirical exercise; and (ii) the interpretation of the selection parameters $(\beta_{p1}, \beta_{p2})$. \\

\noindent
\textit{Economics experiments.---} Consider these three steps in \citet{Camerer2016}:

\begin{enumerate}
    \item \textbf{Eligibility}: Between-study laboratory experiments in \textit{American Economic Review} and \textit{Quarterly Journal of Economics} published between 2011 and 2014.
    \item \textbf{Study selection:} \citet{Camerer2016} select for publication all eligible studies that had `at least one significant between subject treatment effect that was referred to as statistically significant in the paper.' \citet{Andrews2019} review eligible studies and conclude that no studies were excluded by this restriction. Thus, the complete set of eligible studies was selected for replication. 
    \item \textbf{Within-study replication selection:} the most important \textit{statistically significant} result within a study, as emphasized by the authors, was chosen for replication. Further details are in the supplementary materials in \citet{Camerer2016}. Of the 18 replication studies, 16 were significant at the 5\% level and two had $p$-values slightly above 0.05 but were treated as `positive' results for replication and included in the replication rate calculation. 
\end{enumerate}

This selection mechanism implies that the empirical results are valid for the population of `most important' significant results, as emphasized by authors, in experimental economics papers published in top economics journals between 2011 and 2014. Interpretation of the selection parameters is provided in the main text. \\ 

\noindent
\textit{Psychology.---} Next, consider replication selection in \citet{OpenScience2015}:

\begin{enumerate}
    \item \textbf{Eligibility}: Studies published in 2008 in one of the following journals: \textit{Psychological Science}, \textit{Journal of Personality and Social Psychology}, and \textit{Journal of Experimental Psychology: Learning, Memory, and Cognition}.
    \item \textbf{Study selection:} \citet{OpenScience2015} write: ‘The first replication teams could select from a pool of the first 20 articles from each journal, starting with the first article published in the first 2008 issue. Project coordinators facilitated matching articles with replication teams by interests and expertise until the remaining articles were difficult to match. If there were still interested teams, then another 10 articles from one or more of the three journals were made available from the sampling frame.’ Importantly, the most common reason why an article was not matched was due to feasibility constraints (e.g. time, resources, instrumentation, dependence on historical events, or hard-to-access samples). 
    \item \textbf{Within-study replication selection:} the last experiment reported in each article was chosen for replication. \citet{OpenScience2015} write that, `Deviations from selecting the last experiment were made occasionally on the basis of feasibility or recommendations of the original authors.'
\end{enumerate}

This selection mechanism implies that the empirical results are valid for the distribution of last experiments in the set of eligible journals. Since neither studies nor results were selected based on statistical significance, it is reasonable to treat the `last experiment' rule as effectively random. In this case, we can interpret the results are being valid for all results in the eligible set of journals. 

\section*{F. Simulated Replication Rates Under Alternative Power Calculations}

This appendix presents several extensions to the main results on the simulated replication rate. First, it shows simulated replication rate predictions based on alternative rules for calculating power. Second, it calculates the `simulated regression to the mean ratio', a statistic summarizing the extent to which selected samples of significant original estimates are biased. \\

\textit{Alternative power calculation rules.---} I examine two additional `rules' for calculating replication power. For concreteness, suppose we want to calculate the replication standard error for a simulated original study $(x^{sim}, \sigma^{sim}, \theta^{sim})$.  

\begin{enumerate}
    \item \textbf{Common power rule (mean):} This is the rule reported in the results in the main text. It assumes no variability in the application of the common power rule, such that all replications have mean intended power $1-\beta^n$. This rule implies
    \begin{equation}
        \sigma_r^{sim}(x^{sim}, \beta^n) = \frac{|x^{sim}|}{1.96-\Phi^{-1}(\beta^n)}
    \end{equation}
        \item \textbf{Common power rule (realized):} To capture variability in the application of the common power rule, take a random draw from the empirical distribution of $|x|/\sigma_r$ and denote it $1.96 - \widehat{\beta}^n$. Then realized intended power for simulated study $(x^{sim}, \sigma^{sim}, \theta^{sim})$ is equal to
        \begin{equation}
        \sigma_r^{sim}(x^{sim}, \widehat{\beta}^n) = \frac{|x^{sim}|}{1.96-\Phi^{-1}(\widehat{\beta}^n)}
    \end{equation}
    
    Intended power for individual replications varied around mean intended power for at least two reasons. First, replication teams were instructed to meet minimum levels of statistical power, and encouraged to obtain higher power if feasible. Second, a number of replication in \citet{OpenScience2015} did not meet this requirement. Figure \ref{fig:hist_power} shows the distribution of realized intended power in replications for experimental economics and psychology. Realized intended power is right-skewed for psychology. In experimental economics, realized intended power is distributed more tightly around mean. 
    
    \begin{figure} [H]
    	\centering
        \includegraphics[width=0.7\textwidth]{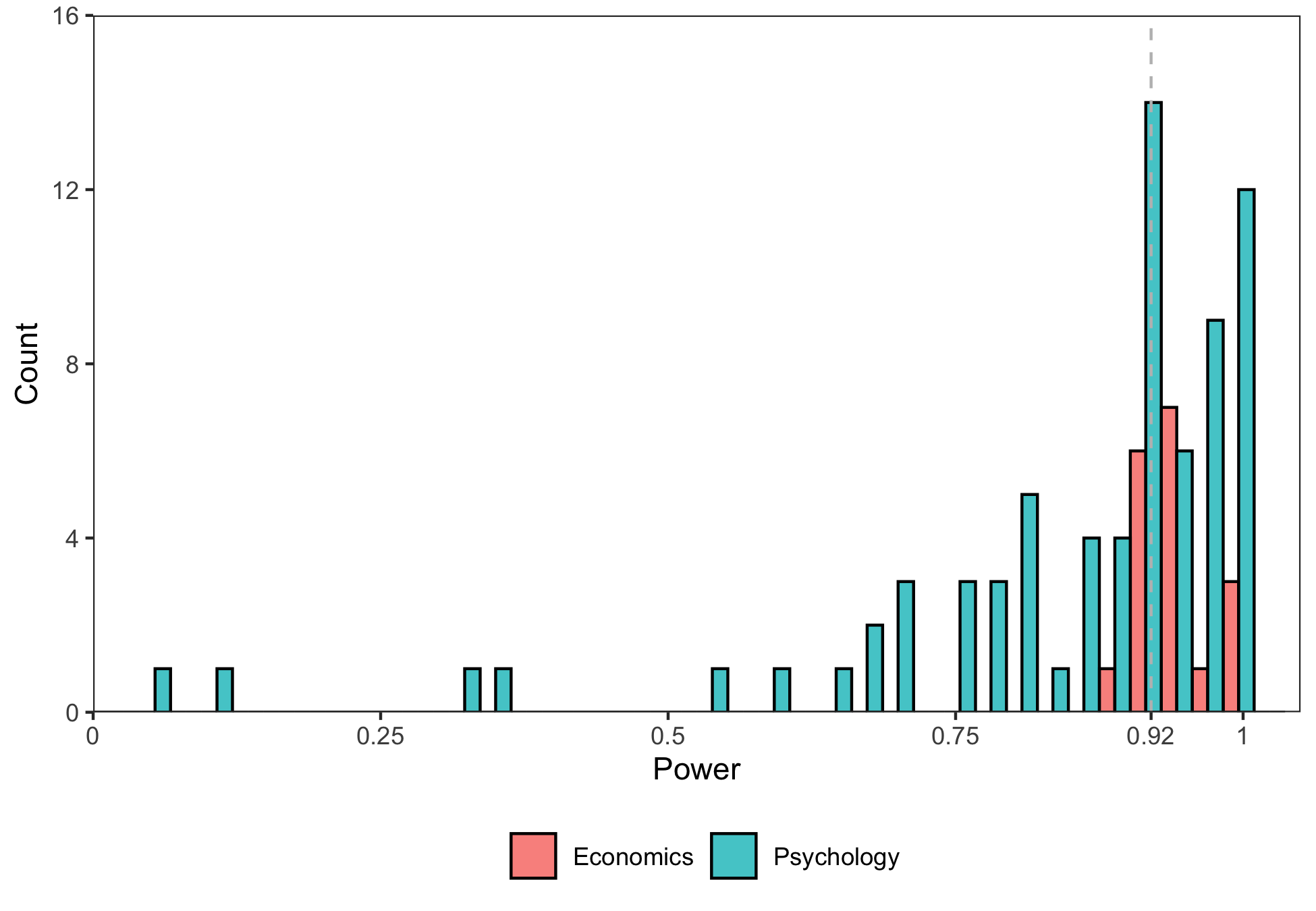} 
    	\caption{Histogram of realized intended power in replication studies in economics experiments and psychology. Data are from \citet{Camerer2016} and \citet{OpenScience2015}, respectively. Realized intended power is defined as $1-\Phi(1.96-\frac{x}{\sigma_r})$. The horizontal dashed line is reported mean power (for both replication studies) of 92\%.}.
    	\label{fig:hist_power}
    \end{figure}

    \item \textbf{Original power:} Set replication power equal to the power in the original study:
    \begin{equation}
        \sigma_r^{sim}(\sigma^{sim}) = \sigma^{sim}
    \end{equation}
    
    This rule has been proposed as a straightforward, intuitive approach for designing replication studies \citep{Anderson2017}.
\end{enumerate}

\textit{Simulated regression to the mean ratio.---} In addition to the simulated replication rate, I report the ratio of mean replication effects to mean original effects among significant results -- referred to as the `regression to the mean ratio'. For any given power rule, let $\{x_i, \sigma_i, x_{r,i}, \sigma_{r,i} \}_{i=1}^{M_{sig}}$ be the set of simulated original studies that are published and significant, and corresponding replication results; $M_{sig}$ is the size of the set. The `simulated regression to the mean ratio' equals

\begin{equation}
    \frac{\frac{1}{M_{sig}} \sum_{i} \tanh{x_{r,i}}}{
    \frac{1}{M_{sig}} \sum_{i} \tanh{x_i}}
    \label{eq:mcrmr}
\end{equation}

\noindent
where the hyperbolic tangent transformation converts the effect size $x$ from Fisher-transformed correlation coefficient units (necessary for estimation and inference) into Pearson correlation coefficient units.\footnote{In both applications, replicators transformed original effect sizes to Pearson correlation coefficients $r$. Standard errors are calculated by applying a Fisher transformation to the correlation coefficients ($z = \text{artanh}(r)$), which is approximately normally distributed \citep{Fisher1915}.} \\

\textit{Results.---}  Table \ref{tab:sim_results_appendix} presents the results. Consider first the results for experimental economics in Panel A. Allowing intended power to vary across replications yields a similar replication rate prediction to assuming all replications have intended power equal to the report mean of 92\%. Setting power in replications equal to power in original studies implies a replication rate of 55.2\%. The second row shows the regression to the mean ratio. The observed regression to the mean ratio in \citet{Camerer2016} is around 0.71. Recall that regression to the mean in replications is induced by the conditioning strategy of the replication rate, and not a consequence of publication bias against null results (Proposition 2). The regression to the mean ratio varies little with power calculations used in the replications.\footnote{Slight variation across different power rules arises because the the hyperbolic tangent transformation to convert Fisher correlation coefficient units back into correlation coefficient units restricts values to be between -1 and 1.}

Panel B shows the results for psychology. Power calculations based on realized intended power gives a slightly lower predicted replication rate than using mean power, because the distribution of intended power is right skewed. Setting power in replications to equal that of original studies leads to a lower predicted replication rate.  The predicted regression to the mean ratios under both common power rules are higher than the observed 0.433.

\begin{table}[H] \centering 
  \footnotesize
  \caption{\textsc{Simulated Replication Rate and Regression to the Mean Ratio Predictions}}
\begin{tabularx}{0.8942\textwidth}{@{\extracolsep{5pt}} lcccc} 
\\[-1.8ex]\hline 
\hline \\[-1.8ex] 
 & &  \multicolumn{3}{c}{\textit{Simulated statistics}} \\ 
 & & \\\cline{3-5}
\textbf{A Economics experiments} & Observed  & $\text{ }$92\% for $X$ &  $\text{ }$Realized power & Original power  \\ 
\hline \\[-1.8ex] 
Replication rate  & 0.611 & 0.601  & 0.625 & 0.552  \\ 
Regression to the mean ratio & 0.617 & 0.709  & 0.710 & 0.704   \\ 
& & & & \\

\textbf{B Psychology experiments} &  & & &     \\ 
\hline \\[-1.8ex] 
Replication rate  & 0.356 & 0.539 & 0.514 & 0.478  \\ 
Regression to the mean ratio & 0.433 & 0.686 & 0.674 & 0.684  \\ 
\hline
\hline
\end{tabularx} 
\caption*{\textit{Notes}: Economics experiments refer to \citet{Camerer2016} and psychology experiments to \citet{OpenScience2015}. The replication rate is defined as the share of original estimate whose replications have statistically significant findings of the same sign. The regression to the mean ratio is defined as the ratio of mean replication effect sizes to mean original effect sizes among significant studies. Figures in the first column are observed outcomes from \citet{Camerer2016} and \citet{OpenScience2015}. Remaining columns report simulated statistics using parameter estimates Table \ref{tab:mle_results} and assuming different rules for calculating power in replications.}
\label{tab:sim_results_appendix}
\end{table} 

\end{document}